\documentclass[journal]{IEEEtran} 

\usepackage{fancyhdr}

\usepackage[T1]{fontenc}
\usepackage{cite}
\usepackage{amsmath,graphicx}
\usepackage{amssymb}
\usepackage{algpseudocode}
\usepackage{amsfonts}
\usepackage{graphicx,subfigure}
\usepackage{fancyhdr}  %
\usepackage{cases}
\usepackage{extarrows}
\usepackage{algorithm}
\usepackage{multirow,tabularx}
\usepackage{mathtools}
\usepackage{xcolor}
\usepackage[english]{babel}
\usepackage{caption}
\usepackage{autobreak}
\usepackage{bm}
\usepackage{graphicx}

\newtheorem{lemma}{Lemma}

\ifCLASSINFOpdf
\else
\fi
\begin{document}
\title{Channel Estimation for RIS-Empowered Multi-User MISO Wireless Communications}
\author{Li Wei, Chongwen Huang, George~C.~Alexandropoulos,~\IEEEmembership{Senior Member,~IEEE,} Chau~Yuen,~\IEEEmembership{Fellow,~IEEE}, Zhaoyang Zhang,~\IEEEmembership{Member,~IEEE}, and M\'{e}rouane~Debbah,~\IEEEmembership{Fellow,~IEEE}
	\thanks{This work was presented in part at the \textit{IEEE SAM}, Hangzhou, China, 8--11 June 2020 \cite{parafac_SAM2020}. (\textit{Corresponding author: Chongwen  Huang}).}
	\thanks{L. Wei and C. Yuen are with the Engineering Product Development (EPD) Pillar, Singapore University of Technology and Design, Singapore 487372 (e-mails: wei\_li@mymail.sutd.edu.sg,  yuenchau@sutd.edu.sg).}
	
	\thanks{C.~Huang and Z.~Zhang are with the College of Information Science and Electronic Engineering, Zhejiang University, Hangzhou 310007, China, and Zhejiang Provincial Key Lab of Information Processing, Communication and Networking (IPCAN), Hangzhou 310007, China, and the International Joint Innovation Center, Zhejiang University, Haining 314400, China (e-mails: \{ning\_ming, chongwenhuang\}@zju.edu.cn).  }
	\thanks{G.~C.~Alexandropoulos is with the Department of Informatics and Telecommunications, National and Kapodistrian University of Athens, Panepistimiopolis Ilissia, 15784 Athens, Greece (e-mail: alexandg@di.uoa.gr). }
	\thanks{M.~Debbah is with CentraleSup\'elec, University Paris-Saclay, 91192 Gif-sur-Yvette, France. M. Debbah is also with the Lagrange Mathematics and Computing Research Center, Paris, 75007 France (email: merouane.debbah@huawei.com). }
}

\maketitle
\thispagestyle{fancy}

\begin{abstract}
Reconfigurable Intelligent Surfaces (RISs) have been recently considered as an energy-efficient solution for future wireless networks due to their fast and low-power configuration, which has increased potential in enabling massive connectivity and low-latency communications. Accurate and low-overhead channel estimation in RIS-based systems is one of the most critical challenges due to the usually large number of RIS unit elements and their distinctive hardware constraints. In this paper, we focus on the uplink of a RIS-empowered multi-user Multiple Input Single Output (MISO) uplink communication systems and propose a channel estimation framework based on the parallel factor decomposition to unfold the resulting cascaded channel model. We present two iterative estimation algorithms for the channels between the base station and RIS, as well as the channels between RIS and users. One is based on alternating least squares (ALS), while the other uses vector approximate message passing to iteratively reconstruct two unknown channels from the estimated vectors. To theoretically assess the performance of the ALS-based algorithm, we derived its estimation Cram\'er-Rao Bound (CRB). We also discuss the downlink achievable sum rate computation with estimated channels and different precoding schemes for the base station. Our extensive simulation results show that our algorithms outperform benchmark schemes and that the ALS technique achieves the CRB. It is also demonstrated that the sum rate using the estimated channels always reach that of perfect channels under various settings, thus, verifying the effectiveness and robustness of the proposed estimation algorithms.
\end{abstract}
	
\begin{IEEEkeywords}
Channel estimation, reconfigurable intelligent surfaces, multi-user communication, parallel factor decomposition, approximate message passing, Cram\'er-Rao bound, sum rate performance.
	\end{IEEEkeywords}

\IEEEpeerreviewmaketitle

\section{Introduction}\label{sec:intro}
\IEEEPARstart{R}{econfigurable} Intelligent Surfaces (RISs) and metasurfaces are lately gaining increasing interest as a candidate technology for beyond 5-th Generation (5G) wireless communication, mainly due to their significant potential for enabling low-power, energy-efficient, high-speed, massive-connectivity, and low-latency wireless communications\cite{huang2020holographic,holobeamforming,husha_LIS2,liaskos2018new,huang2019reconfigurable,Marco2019,qingqing2019towards,WavePropTCCN,9206044}. They are artificial planar structures with integrated electronic circuits that can be programmed to manipulate an incoming electromagnetic field in a customizable way. An RIS usually consists of a large number of hardware-efficient and nearly passive reflecting elements, each of which can alter the phase of the incoming signal without requiring a dedicated power amplifier, as for example needed in conventional Amplify-and Forward (AF) relaying systems \cite{holobeamforming,huang2019reconfigurable}.
	
The energy efficiency potential of RIS in the uplink of outdoor multi-user Multiple Input Single Output (MISO) uplink communications were analyzed in \cite{huang2019reconfigurable} and \cite{9200578}, while \cite{husha_LIS2} focused on an indoor scenario to illustrate the potential of RIS-based positioning. It was shown in \cite{huang2018energy,han2019} that the latter gains are large even in the case of RISs with finite resolution unit elements and statistical channel knowledge. Recently, \cite{8941126} proposed a novel passive beamforming and information transfer technique to enhance primary communication, and a two-step approach at the receiver to retrieve the information from both the transmitter and RIS. Orthogonal and non-orthogonal multiple access in RIS-assisted communications was studied in \cite{9133094} as a cost-effective solution for boosting spectrum/energy efficiency. In physical-layer security systems, artificial noise \cite{8972400} was lately combined with RISs to enhance the secrecy rate performance \cite{9134962, RIS_PLS_2020,wanghuimingSecure}.    RIS-assisted communications were recently investigated in the millimeter wave (mmWave) and Terahertz (THz)  bands, where RIS was employed to solve the limited transmission distance problem  \cite{Akyildiz2018mag}. The existing research works have proved the great potential of RIS, however, most of the existing research works focusing on RIS configuration optimization techniques assumed the perfect channel state information availability. 

Channel Estimation (CE) in RIS-empowered multi-user communications is a challenging task since it involves the estimation of multiple channels simultaneously: the direct channels between the Base Station (BS) and each user, the channels between the RIS and BS, and the channels between the RIS and each user \cite{nadeem_ce,9133156,9144510}. This task becomes more complex when the deployed RISs are equipped with large numbers of unit elements having non-linear hardware characteristics \cite{huang2020holographic,9324910}. In \cite{8879620}, a general framework for cascaded CE in RIS-assisted Multiple Input Multiple Output (MIMO) systems was introduced by leveraging combined bilinear sparse matrix factorization and matrix completion. A control protocol enabling linear square estimation for the involved channel vectors in the MISO case was presented in \cite{8683663}. According to this protocol, some of the RIS elements were activated during the CE phase. The authors in \cite{9053695} claimed that they designed an optimal CE scheme, where the RIS unit elements follow an optimal series of activation patterns.   In \cite{9081935}, the authors designed the joint optimal training sequence and reflection pattern to minimize the channel estimation MSE for the RIS-aided wireless system by recasting the corresponding problem into the convex semi-definite programming problem, however, the authors just estimate cascaded channels  instead of separate channels in the RIS-based system.      In \cite{9054415}, an RIS-based activity detection and CE problem was formulated as a sparse matrix factorization, matrix completion, and multiple measurement vector problem that was solved via the approximate message passing algorithm. In \cite{9104260}, authors proposed a closed-form least squares Khatri-Rao factorization (LSKRF) estimation algorithm for RIS-assisted MIMO systems, which has a low complexity but a more constrictive requirement on the training parameters. Very recently \cite{8937491}, a transmission protocol for CE and RIS configuration optimization was proposed for RIS-enhanced orthogonal frequency division multiplexing systems. However, the presented RIS reflection pattern implements ON/OFF switching, which can be costly in practice, and the CE accuracy degrades due to the fact that only a  portion of RISs are activated during the CE.

To lower the protocol overhead for CE in RIS-empowered systems, machine learning approaches have been lately presented\cite{huang2019spawc,Alkhateeb2019,hardware2020icassp,chongwendrl}. Considering an indoor scenario in \cite{huang2019spawc}, a deep neural network was designed to unveil the mapping between the measured coordinate information at a user location and the configuration of the RIS unit elements that maximizes this user's received signal strength. In \cite{Alkhateeb2019}, the authors considered RISs with multiple active elements for partial channels sensing at their sides, and presented a compressive sensing based solution to recover the full channels from the sampled ones. Very recently \cite{hardware2020icassp}, an RIS architecture consisting of any number of passive reflecting elements and a single radio frequency chain for baseband measurements was presented that was deployed for the estimation of sparse channels at the RIS side via matrix completion tools. However, the machine learning methods \cite{huang2019spawc,Alkhateeb2019,hardware2020icassp} either require extensive training during offline phases or assume that RISs have some active elements realizing analog or hybrid analog and digital reception. Inevitably, the latter architectures increase the RIS hardware complexity and power consumption.

Recently, the PARAllel FACtor (PARAFAC) decomposition \cite{harshman1994parafac,7556337,bro2003new,ten2002uniqueness,roemer2008closed} has been  considered as an efficient method for estimating multiple large channel matrices in MIMO communication systems \cite{852018,du2014low}. It enables the decomposition of a high dimensional tensor (or matrix) into a linear combination of multiple rank-one tensors, thus, facilitating low complexity estimation of low-rank matrices. In \cite{rong2012channel}, a PARAFAC-based CE approach for two-hop MIMO relay systems was presented that required less number of training data blocks. The authors in \cite{ximenes2014parafac} applied the PARAFAC decomposition for CE in a two-hop AF MIMO relay system, alleviating the need for estimation at the relay node. Inspired by PARAFAC's promising results in CE for relay systems, we proposed very recently in \cite{parafac_SAM2020} an Alternating Least Squares (ALS) CE algorithm for RIS-assisted multi-user MISO uplink communications, which exhibited the promising performance gains compared to state-of-the-art schemes. However, the performance of the presented algorithm was not theoretically studied, and hence, the accuracy and stability of the deployed PARAFAC decomposition were not thoroughly investigated.
	
In this paper, motivated by the PARAFAC decomposition \cite{rong2012channel, ximenes2014parafac}, we leverage ALS and the Vector Approximate Message Passing (VAMP) algorithm \cite{8713501} to iteratively recover signals from noisy observations. We consider the uplink of a RIS-empowered multi-user MISO communication system and present two estimation algorithms for the channel matrices between the BS and RIS as well as the multiple users and the RIS. The presented algorithms capitalize on the unfolded forms \cite{sidiropoulos2000blind,kolda2009tensor} of the involved matrices to yield efficient CE. Our representative simulation results validate the accuracy of the proposed techniques and their favorable performance over state-of-the-art schemes, as well as their impact on the downlink sum rate performance when compared with the perfect CE case. The main contributions of this paper are summarized as follows:
	\begin{itemize}
		\item  We introduce a PARAFAC-based CE framework for RIS-empowered communications enabling the efficient estimation of multiple large channel matrices. The high dimensional tensor that involves the unknown channels is represented in different unfolded forms through the PARAFAC decomposition. Leveraging those forms, we adopt ALS and VAMP to estimate the unknown channel matrices.
		\item We present the feasibility conditions and the computational complexity for both the proposed CE algorithms. In order to theoretically assess the performance of the ALS-based algorithm, we derived the  Cram\'er-Rao Bound (CRB) of its estimation performance. 
		\item We investigate the achievable downlink sum rate of the considered RIS-empowered system with different BS precoding schemes, namely, the Maximum Ratio Transmission (MRT), Minimum Mean Square Error (MMSE), and Zero Forcing (ZF) schemes, using the estimated channels from both designed algorithms.
		\item We prove the validity of the proposed algorithms through simulation results by comparing  with state-of-the art techniques and the perfect CE case. It also is shown that the proposed ALS algorithm can achieve the CRB in various cases.
	\end{itemize}

The remainder of this paper is organized as follows. In Section \ref{sec:format}, the considered signal and system models are included together with the problem formulation. Our PARAFAC-based CE algorithms are presented in Section~\ref{sec:channel_est}, whereas Section~\ref{sec:crb} derives the CRB for the proposed ALS-based technique. Section~\ref{sec:sum_rate} discussed the achievable sum rate performance computation with different precoding schemes and CE. Finally, concluded  remarks are drawn in Section~\ref{sec:conclusion}.
	
\textit{Notation}: Fonts $a$, $\mathbf{a}$, and $\mathbf{A}$ represent scalars, vectors, and matrices, respectively. $\mathbf{A}^T$, $\mathbf{A}^H$, $\mathbf{A}^{-1}$, $\mathbf{A^\dag}$, and $\|\mathbf{A}\|_F$ denote transpose, Hermitian (conjugate transpose), inverse, pseudo-inverse, and Frobenius norm of $ \mathbf{A} $, respectively. $[\mathbf{A}]_{i,j}$ represents $\mathbf{A}$'s $(i,j)$-th element, while $[\mathbf{A}]_{i,:}$ and $[\mathbf{A}]_{:,j}$ stand for its $i$-th row and $j$-th column, respectively. $|\cdot|$ and $(\cdot)^*$ denote the modulus and conjugate, respectively. $\text{tr}(\cdot)$ gives the trace of a matrix, $\mathbf{I}_n$ (with $n\geq2$) is the $n\times n$ identity matrix, and $\mathbf{1}_n$ is a column vector with all ones. The rank of $\mathbf{A}$ is denoted as ${\text{rank}}(\mathbf{A})$ and $k_{\mathbf{A}}\leq{\text{rank}}(\mathbf{A})$ stands for $\mathbf{A}$'s Kruskal rank (or $k$-rank), defined as the maximum integer $k$ such that any $k$ columns drawn from $\mathbf{A}$ are linearly independent. $\otimes$ and $\circ$ represent the Kronecker and the Khatri-Rao (column wise Kronecker) matrix products, respectively. $\delta_{k,i}$ equals to $1$ when $k=i$ or $0$ when $k\neq i$, and $\mathbf{e}_n$ is the $n$-th unit coordinate vector with $1$ in the $n$-th basis and $0$'s in each $n^\prime$-th basis $\forall$$n^\prime \neq n$. Finally, notation ${\rm diag}(\mathbf{a})$ represents a diagonal matrix with the entries of $\mathbf{a}$ on its main diagonal, and SVD stands for the Singular Value Decomposition. $\delta(\cdot)$ is the Dirac delta function.

\section{System and Signal Models}\label{sec:format}
In this section, we first describe the signal and system models for the considered RIS-empowered multi-user MISO uplink communication system, and then present the PARAFAC decomposition for the end-to-end wireless communication channel.
\subsection{System Model}\label{subsec:signal model}
\begin{figure}\vspace{-4mm}
		\begin{center}
			\centerline{\includegraphics[width=0.47\textwidth]{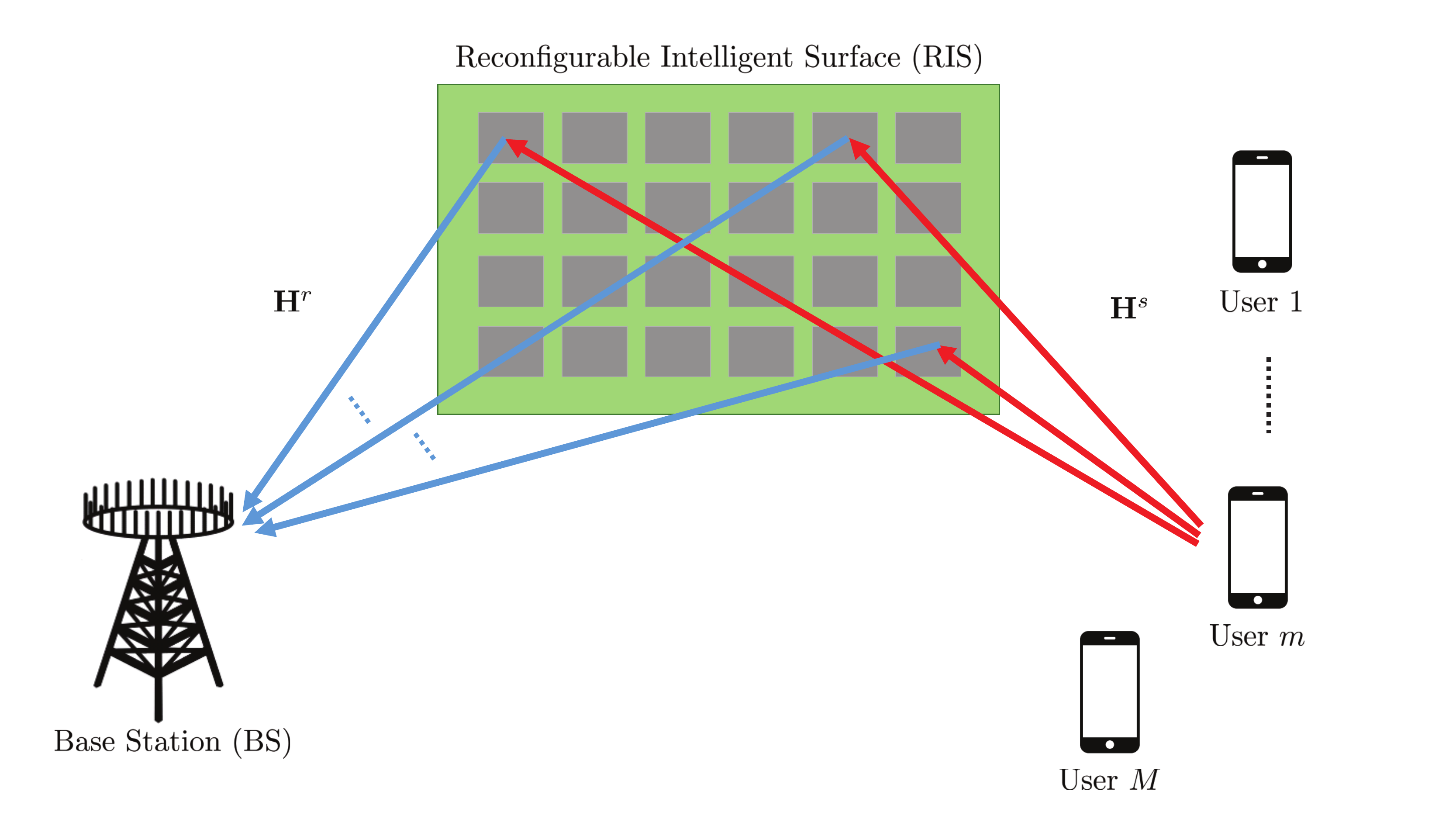}}  \vspace{-0mm}
			\caption{The considered RIS-empowered multi-user MISO communication system consisting of an $K$-antenna BS and $M$ single-antenna mobile users.  }
			\label{fig:Estimation_Scheme}  
		\end{center}
	\end{figure}

Consider the uplink communication between a BS equipped with $K$ antenna elements and $M$ single-antenna mobile users. We assume that this communication is realized via a passive RIS with discrete unit elements \cite{huang2020holographic}, which is deployed on the facade of a building existing in the vicinity of the BS side, as illustrated in Fig$.$~\ref{fig:Estimation_Scheme}.   The direct channel paths between the BS and the mobile users are assumed highly attenuated due to unfavorable propagation conditions, e.g., the mmWave signal may be greatly attenuated during the direct propagation. Thus, we have this assumption that the direct link can be ignored for the considered RIS-empowered system. In addition, our proposed method also works in the scenario where there exists the direct link. Specifically, the direct link can be estimated easily by traditional methods through the setting that all RIS are powered off.  

The RIS is comprised of $N$ unit cells of equal small size, each made from metamaterials that are capable of adjusting their reflection coefficients. The received discrete-time signals at all $K$ antennas for $T$ consecutive time slots using the $p$-th feasible RIS phase configuration (out of the $P$ available in total; hence, $p=1,2,\ldots,P$) can be compactly expressed with the matrix $\mathbf{Y}_{p}\in\mathbb{C}^{K \times T}$ given by
\begin{equation}\label{equ:YXZ}
\mathbf{Y}_{p}\triangleq\mathbf{H}^{r} D_p(\mathbf{\Phi})  \mathbf{H}^{s} \mathbf{X}+\mathbf{W}_{p},
\end{equation}
where $D_p(\mathbf{\Phi})\triangleq {\rm diag} ([\mathbf{\Phi}]_{p,:})$ with $[\mathbf{\Phi}]_{p,:}$ representing the $p$-th row of the $P\times N$ complex-valued matrix $\mathbf{\Phi}$, which includes at its $P$ rows all feasible RIS phase configurations that are usually chosen from low resolution discrete sets \cite{huang2018energy}. Notations $\mathbf{H}^{r}\in\mathbb{C}^{K\times N}$ and $\mathbf{H}^{s}\in\mathbb{C}^{N\times M}$  represent the channel matrices between RIS and BS and between all $M$ users and RIS, respectively. Additionally, $\mathbf{X}\in\mathbb{C}^{M \times T}$ includes the transmitted signals within the $T$ time slots; it must hold $T\geq M$ for efficient CE. Finally, $\mathbf {W}_p \in \mathbb{C}^{K \times T}$ denotes the Additive White Gaussian Noise (AWGN) matrix having elements with zero mean and variance $\sigma^2$.
	
\subsection{Preliminaries on the PARAFAC Decomposition}\label{subsec:parafac}
The PARAFAC decomposition is a subset of multi-way analysis inspired from psychometrics and chemometrics, and has been widely adopted in analyzing high-dimensional data \cite{sidiropoulos2000blind}. Let the matrices (also known as two-way arrays) $\mathbf{A}\in \mathbb{C}^{I \times R},\mathbf{B}\in \mathbb{C}^{J \times R}$ and $\mathbf{C}\in \mathbb{C}^{K \times R}$ having elements $[\mathbf{A}]_{i,r}$, $[\mathbf{B}]_{j,r}$ and $[\mathbf{C}]_{k,r}$, respectively, with $i=1,2,\ldots,I$, $j=1,2\ldots,J$, $k=1,2,\ldots,K$, and $r=1,2,\ldots,R$. We construct the $(i,j,k)$-th element of the three-way (i.e., three dimensional) array $\mathbf{X}\in \mathbb{C}^{I \times J \times K}$ from the two-way arrays $\mathbf{A}$, $\mathbf{B}$, and $\mathbf{C}$ 
as follows:
\begin{equation} \label{equ:parafc_scalar}
	[\mathbf{X}]_{i,j,k} \triangleq \sum^{R}_{r=1} [\mathbf{A}]_{i,r} [\mathbf{B}]_{j,r} [\mathbf{C}]_{k,r}.
\end{equation}
It can be easily shown that the two-way slices (i.e., matrix slices) of $\mathbf{X}$ can be expressed as
	 \begin{equation} \label{equ:parafac}
	 \begin{array}{llll}
	 \mathbf{X}_{i,:,:} & = & \mathbf{B} \operatorname{diag}\left([\mathbf{A}]_{i,:}\right)\mathbf{C}^T, & i=1,2, \ldots, I, \\
	 \mathbf{X}_{:,j,:} & = & \mathbf{C} \operatorname{diag}\left([\mathbf{B}]_{j,:}\right)\mathbf{A}^T, & j=1,2, \ldots, J, \\
	 \mathbf{X}_{:,:,k} & = & \mathbf{A} \operatorname{diag}\left([\mathbf{C}]_{k,:}\right) \mathbf{B}^T, & k=1,2,\ldots, K.
	 \end{array}\end{equation}
This implies that the symmetry in \eqref{equ:parafc_scalar}	leads to different matrix rearrangements, which enables alternative views of $\mathbf{X}$. In the sequel, we will exploit this PARAFAC decomposition to devise two CE techniques for the considered RIS-empowered multi-user MISO system.

\subsection{Received Training Symbols}\label{subsec:prob_formu}	
The channel matrices $\mathbf{H}^{s}$ and $\mathbf{H}^{r}$ in \eqref{fig:Estimation_Scheme} are in general unknown and needed to be estimated. We hereinafter assume that these matrices have independent and identically distributed complex Gaussian entries (i.e., Rayleigh fading envelopes); the entries between these two matrices are also assumed independent. We further assume orthogonal pilot signals in \eqref{fig:Estimation_Scheme}, such that $\mathbf{X} \mathbf{X}^{H}=\mathbf{I}_{M}$, and that $\mathbf{\Phi}$ with the $P$ distinct configurations is used during the CE phase. In order to estimate $\mathbf{H}^{s}$ and $\mathbf{H}^{r}$, we first apply the PARAFAC decomposition in \eqref{equ:YXZ}, according to which the unknown wireless channels are represented using tensors \cite{kroonenberg1980principal,rong2012channel}. In doing so, we define for each $p$-th RIS training configuration the matrix $\widetilde{\mathbf{Z}}_{p}\in\mathbb{C}^{K\times M}$, after removing the pilot symbols from each received training signal $\mathbf{Y}_{p}$, as
	\begin{equation}\label{widetilde_Z_p}
	\widetilde{\mathbf{Z}}_{p}\triangleq\mathbf{Y}_{p}\mathbf{X}^{H}=\underbrace{\mathbf{H}^{r} D_p(\mathbf{\Phi})  \mathbf{H}^{s}}_{\triangleq \mathbf{Z}_{p}}+\widetilde{\mathbf{W}}_{p},
	\end{equation}
where $\mathbf{Z}_{p} \in \mathbb{C}^{K \times M}$ is the noiseless version of the end-to-end RIS-based wireless channel and $\widetilde{\mathbf{W}}_{p}\triangleq\mathbf{W}_{p}\mathbf{X}^{H}\in \mathbb{C}^{K \times M}$ is the noise matrix after removing the pilot symbols. It can be easily shown that each $(k,m)$-th entry of $\mathbf{Z}_{p}$ with $k=1,2,\ldots,K$ and $m=1,2,\ldots,M$ is obtained as
\begin{equation}\label{scalar_decomp}
[\mathbf{Z}_{p}]_{k,m}=\sum_{n=1}^{N} [\mathbf{H}^{r}]_{k,n} [\mathbf{H}^{s}]_{n,m} [\mathbf{\Phi}]_{p,n}.
\end{equation}

Each $\mathbf{Z}_{p}$ in \eqref{scalar_decomp}, out of the $P$ in total, can be equivalently represented with three distinct matrix forms, as follows. We formulate the three-way matrix $\mathbf{Z}\in \mathbb{C}^{K \times M \times P}$ that includes all $P$ matrices $\mathbf{Z}_p$ in \eqref{widetilde_Z_p} in its third dimension. Then, the unfolded forms \cite{sidiropoulos2000blind,kolda2009tensor} of the mode-1, mode-2, and mode-3 of $\mathbf{Z}$ can be expressed as
	\begin{align}
	\label{mode_1}\text{Mode-1:}& \quad {\mathbf{Z}^1\triangleq(\left(\mathbf{H}^{s}\right)^{T} \circ   \mathbf{\Phi})\left(\mathbf{H}^{r}\right)^{T}}\in\mathbb{C}^{PM \times K},
	\\
	\label{mode_2}\text{Mode-2:}&  \quad {\mathbf{Z}^2\triangleq( \mathbf{\Phi} \circ \mathbf{H}^r)\mathbf{H}^s}\in\mathbb{C}^{KP \times M},
  \\
	\label{mode_3}\text{Mode-3:}& \quad {\mathbf{Z}^3\triangleq(\mathbf{H}^r \circ \left(\mathbf{H}^{s}\right)^{T})  \mathbf{\Phi}^T}\in\mathbb{C}^{MK \times P}.
	\end{align}
Note that $\mathbf{Z}^1$, $\mathbf{Z}^2$, and $\mathbf{Z}^3$ can be regarded as the horizontal, lateral, and frontal slices of $\mathbf{Z}$. Clearly, each of the postprocessed noisyless received channel matrices $\mathbf{Z}^1$, $\mathbf{Z}^2$, and $\mathbf{Z}^3$ is given as a matrix product of a Khatri-Rao matrix product factor with a single  matrix. It will be shown in the following that this expression facilitates the considered CE problem.
	
\section{Proposed PARAFAC-Based CE}\label{sec:channel_est}
In this section, we capitalize on the previously described PARAFAC decomposition and present two iterative estimation algorithms, namely ALS and VAMP, for all involved channel matrices.   Based on the unfolded forms of the received signal, two separate channels can be estimated iteratively from the matrix slices.    We also discuss the feasibility conditions for both proposed algorithms.

\subsection{ALS Channel Estimation}\label{subsec:als_est}
We let $\widetilde{\mathbf{W}} \in \mathbb{C}^{K \times M \times P}$ include all $P$ AWGN matrices $\widetilde{\mathbf{W}}_p$ in \eqref{widetilde_Z_p} and then formulate the three-way matrix $\widetilde{\mathbf{Z}}\in \mathbb{C}^{K \times M \times P}$ as
	\begin{equation}\label{noise_term}
	\widetilde{\mathbf{Z}}\triangleq\mathbf{Z}+\widetilde{\mathbf{W}}.
	\end{equation}
It is obvious that $\mathbf{Z}$ obtained from \eqref{scalar_decomp} represents the noiseless version of $\widetilde{\mathbf{Z}}$, which can be unfolded according to the forms in \eqref{mode_1}--\eqref{mode_3}. In a similar way, we can get the unfolded forms $\widetilde{\mathbf{Z}}^1\in\mathbb{C}^{PM \times K}$, $\widetilde{\mathbf{Z}}^2\in\mathbb{C}^{KP \times M}$, and $\widetilde{\mathbf{Z}}^3\in\mathbb{C}^{MK \times P}$ of $\widetilde{\mathbf{Z}}$. For example, ${\widetilde{\mathbf{Z}}^1\triangleq(\left(\mathbf{H}^{s}\right)^{T} \circ   \mathbf{\Phi})\left(\mathbf{H}^{r}\right)^{T}} + \widetilde{\mathbf{W}}^1$, where the $PM \times K$ AWGN matrix $\widetilde{\mathbf{W}}^1\in\mathbb{C}^{PM \times K}$ is obtained from $\widetilde{\mathbf{W}}$. Inspired by this PARAFAC decomposition \cite{kroonenberg1980principal}, we next present the algorithmic steps for the proposed ALS-based iterative CE algorithm, which is also summarized in Algorithm~1. It is noted that the designed algorithm needs to be implemented in practice on the mobile users.

\subsubsection{\textbf{Initialization}}
We initiate the proposed iterative algorithm with the eigenvector matrices presented in \cite{kroonenberg1980principal}. In particular, we set as $\widehat{\mathbf{H}}^{s}_{(0)}$ the matrix including the eigenvectors corresponding to the $N$ non-zero eigenvalues of $\left(\widetilde{\mathbf{Z}}^{2}\right)^H \widetilde{\mathbf{Z}}^2$, where $\widetilde{\mathbf{Z}}^2\in \mathbb{C}^{KP \times M}$ is the mode-2 form of \eqref{noise_term}; this is \eqref{mode_2}'s noisy version. Similarly, $\widehat{\mathbf{H}}^{r}_{(0)}$ is the eigenvector matrix corresponding to the $N$ non-zero eigenvalues of $\left(\widetilde{\mathbf{Z}}^{1}\right)^H \widetilde{\mathbf{Z}}^1$ with $\widetilde{\mathbf{Z}}^1\in \mathbb{C}^{PM \times K}$ being the mode-1 form of \eqref{noise_term}, which is actually the noisy version of \eqref{mode_1}. It is noted that our proposed algorithm requires a priori feasible $\mathbf{\Phi}$ configuration. To this end, we design $\mathbf{\Phi}$ as a discrete Fourier transform matrix when we estimate the channels of $\mathbf{H}^s$ and $\mathbf{H}^r$, which is recommended as a good choice for PARAFAC-based CE \cite{rong2012channel}.   Specifically, $\mathbf{\Phi}$ is chosen from the first $P$ rows of $N \times N$ Fourier transform matrix. 
	
\subsubsection{\textbf{Iterative Update}}
The channels $\mathbf{H}^s$ and $\mathbf{H}^r$ are obtained in an iterative manner by alternatively minimizing conditional least squares (LS) criteria using the postprocessed received signal $\widetilde{\mathbf{Z}}$ given by \eqref{noise_term} and the unfolded forms \eqref{mode_1}--\eqref{mode_3} for $\mathbf{Z}$. Starting with $\mathbf{H}^r$, we make use of the mode-1 unfolded form given by \eqref{mode_1} according to which $\mathbf{Z}^1=\mathbf{A}^1\left(\mathbf{H}^{r}\right)^T$ with $\mathbf{A}^1\triangleq\left(\mathbf{H}^{s}\right)^T\circ \mathbf{\Phi} \in \mathbb{C}^{PM \times N}$. At the $i$-th algorithmic iteration, the $i$-th estimation for $\mathbf{H}^r$, denoted by $\widehat{\mathbf{H}}^r_{(i)}$, is obtained from the minimization of the following LS objective function:
	\begin{equation}\label{ls_hr}
	\begin{aligned} J\left(\widehat{\mathbf{H}}^{r}_{(i)}\right)=\left\|\widetilde{\mathbf{Z}}^1-\widehat{\mathbf{A}}^{1}_{(i-1)}\left(\widehat{\mathbf{H}}^{r}_{(i)}\right)^T\right\|_{F}^{2},
	\end{aligned}
	\end{equation}
where $\widetilde{\mathbf{Z}}^1\in\mathbb{C}^{PM \times K}$ is a matrix-stacked form of \eqref{noise_term}'s three-way matrix $\widetilde{\mathbf{Z}}$, and $\widehat{\mathbf{A}}^{1}_{(i-1)}=\left(\widehat{\mathbf{H}}^{s}_{(i-1)}\right)^T\circ\mathbf{\Phi}$. The closed-form solution of \eqref{ls_hr} is given by
\begin{equation}\label{ls_hr_solution}
	\left(\widehat{\mathbf{H}}^{r}_{(i)}\right)^T=\left(\widehat{\mathbf{A}}^{1}_{(i-1)}\right)^{\dagger}\widetilde{\mathbf{Z}}^1.
\end{equation}
	
In a similar way, to estimate $\mathbf{H}^s$, we use the mode-2 unfolded form \eqref{mode_2} according to which ${\mathbf{Z}^2=\mathbf{A}^2\mathbf{H}^s}$ with $\mathbf{A}^2\triangleq\mathbf{\Phi} \circ\mathbf{H}^r \in \mathbb{C}^{KP \times N}$, and formulate the following LS objective function for the $i$-th estimation for $\mathbf{H}^s$ denoted by $\widehat{\mathbf{H}}^{s}_{(i)}$ 
	\begin{equation}\label{ls_hs}
	\begin{aligned} J\left(\widehat{\mathbf{H}}^s_{(i)}\right) &=\left\|\widetilde{\mathbf{Z}}^2-\widehat{\mathbf{A}}^{2}_{(i)}\widehat{\mathbf{H}}^s_{(i)}\right\|_{F}^{2}.
	\end{aligned}
	\end{equation}
In the latter expression, $\widetilde{\mathbf{Z}}^2\in\mathbb{C}^{KP \times M}$ is another matrix-stacked form of $\widetilde{\mathbf{Z}}$ and $\widehat{\mathbf{A}}^{2}_{(i)}= \mathbf{\Phi}\circ\widehat{\mathbf{H}}^{r}_{(i)}$. Problem \eqref{ls_hs} is solved with the following closed form expression 
	\begin{equation}\label{ls_hs_solution}
	\begin{aligned}
	\widehat{\mathbf{H}}^{s}_{(i)}=\left(\widehat{\mathbf{A}}^{2}_{(i)}\right)^{\dagger}\widetilde{\mathbf{Z}}^2.
	\end{aligned}
	\end{equation}
	
\subsubsection{\textbf{Iteration Termination Criterion}}
The proposed iterative ALS algorithm terminates when either the maximum number $I_{\rm max}$ of algorithmic iterations is reached or the Normalized Mean Square Error (NMSE) between any adjacent two iterations is less than the threshold $\kappa$ (see the mathematical criterion in the Step $5$ of Algorithm~1). Note that the iterative estimation of $\mathbf{H}^s$ and $\mathbf{H}^r$ using the ALS approach shown in Algorithm~1 encounters a scaling ambiguity from the convergence point, which can be resolved with adequate normalization \cite{rong2012channel}. 
 
\begin{algorithm}[!t]\label{alg:als}\caption{\textbf{Proposed Iterative CE Based on ALS}}
\begin{algorithmic}[1]
  \Require A feasible $\mathbf{\Phi}$, $\kappa>0$, and the number of maximum algorithmic iterations $I_{\rm max}$.
	\State \textbf{Initialization:} $\widehat{\mathbf{H}}^r_{(0)}$, $\widehat{\mathbf{H}}^s_{(0)}$, and set the algorithmic iteration as $i=1$.
	\For {$i=1,2,\ldots,I_{\rm max}$}
	\State Set $\widehat{\mathbf{A}}^{1}_{(i-1)}=\left(\widehat{\mathbf{H}}^{s}_{(i-1)}\right)^T\circ\mathbf{\Phi}$ and then compute $\widehat{\mathbf{H}}^r_{(i)}=\left(\left(\widehat{\mathbf{A}}^{1}_{(i-1)}\right)^{\dagger}\widetilde{\mathbf{Z}}^1\right)^T$.
	\State Set $\widehat{\mathbf{A}}^{2}_{(i)}= \mathbf{\Phi}\circ\widehat{\mathbf{H}}^{r}_{(i)}$ and compute $\widehat{\mathbf{H}}^s_{(i)}=\left(\widehat{\mathbf{A}}^{2}_{(i)}\right)^{\dagger}\widetilde{\mathbf{Z}}^2$.
	\State \textbf{Until} {$\|\widehat{\mathbf{H}}^{r}_{(i)}-\widehat{\mathbf{H}}^{r}_{(i-1)} \|_{F}^{2}\|\widehat{\mathbf{H}}^{r}_{(i)} \|_{F}^{-2} \leq \kappa$ and $\|\widehat{\mathbf{H}}^{s}_{(i)}-\widehat{\mathbf{H}}^{s}_{(i-1)} \|_{F}^{2}\|\widehat{\mathbf{H}}^{s}_{(i)} \|_{F}^{-2} \leq \kappa$} or $i>I_{\rm max}$.
  \EndFor
\Ensure $\widehat{\mathbf{H}}^r_{(i)}$ and $\widehat{\mathbf{H}}^s_{(i)}$ that are the estimations of $\mathbf{H}^r$ and $\mathbf{H}^s$, respectively.
\end{algorithmic}
\end{algorithm}

\subsection{VAMP Channel Estimation}\label{subsec:vamp_est}
An alternative approach to the LS steps in Algorithm 1 can be developed based on the recently proposed VAMP algorithm \cite{8713501,fletcher2017learning}  or the UTAMP algorithm,  which exhibit robustness in recovering signals from noisy measurements \cite{9293406,guo2015approximate}. Motivated by replacing matrix inverse computation, we next present an alternative that adopts VAMP to ALS estimation technique.

Recall that the $PM\times K$ matrix $\widetilde{\mathbf{Z}}^1$ is obtained from the $K\times M\times P$ matrix $\widetilde{\mathbf{Z}}$ in \eqref{noise_term}. We commence by expressing the $k$-th column of $\widetilde{\mathbf{Z}}^1$, with $k=1,2,\ldots,K$, as follows:
\begin{equation}\label{mode1_vamp}
	[\widetilde{\mathbf{Z}}^1]_{:,k} =\mathbf{A}^1[\left(\mathbf{H}^{r}\right)^T]_{:,k}+[\widetilde{\mathbf{W}}^1]_{:,k}\in\mathbb{C}^{PM \times 1},
\end{equation}	
where $\widetilde{\mathbf{W}}^1\in\mathbb{C}^{PM \times K}$ is a matrix-stacked form of the three-way matrix $\widetilde{\mathbf{W}}$ in \eqref{noise_term}, $\widetilde{\mathbf{Z}}^1$ is the observation matrix, and  ${\mathbf{A}}^1$ is the measurement matrix. Similarly, the $m$-th column ($m=1,2,\ldots,M$) of the $KP\times M$ matrix $\widetilde{\mathbf{Z}}^2$ is expressed as
\begin{equation}\label{mode2_vamp}
  [\widetilde{\mathbf{Z}}^2]_{:,m}=\mathbf{A}^2[\mathbf{H}^{s}]_{:,k}+[\widetilde{\mathbf{W}}^2]_{:,m}\in\mathbb{C}^{KP \times 1},
\end{equation}
where $\widetilde{\mathbf{W}}^2\in\mathbb{C}^{KP \times M}$ is another matrix-stacked form of $\widetilde{\mathbf{W}}$ in \eqref{noise_term}, as well as $\widetilde{\mathbf{Z}}^2$ and ${\mathbf{A}}^2$ also denote the observation and measurement matrices, respectively.

The algorithmic steps for the VAMP-based computation of $\hat{\mathbf{H}}^s_{(i)}$ and $\hat{\mathbf{H}}^r_{(i)}$ are provided in Algorithm~\ref{alg:vamp}. Recall that these matrices are computed via LS in the Steps $3$ and $4$ of Algorithm~1. As shown in Algorithm~\ref{alg:vamp}, any of the $\hat{\mathbf{H}}^s_{(i)}$ and $\hat{\mathbf{H}}^r_{(i)}$ is calculated iteratively using \eqref{mode1_vamp} and \eqref{mode2_vamp}. For the simplicity of the presentation of the algorithm, we use the notation $[\widetilde{\mathbf{Z}}]_{:,c}$ with $c=1,2,\ldots,C$ to represent the $c$-th column of $\widetilde{\mathbf{Z}}^1$ when $C=K$ and $\widetilde{\mathbf{Z}}^2$ for $C=M$. In a similar manner, $\widehat{\mathbf{h}}_c$ represents the $c$-th column of the estimation $\left(\mathbf{H}^{r}\right)^T$ or $\mathbf{H}^{s}$. In addition, notation $\widehat {\mathbf{A}}$ refers to $\widehat {\mathbf{A}}^1$ when Algorithm~\ref{alg:vamp} is used to estimate $\left( \mathbf{H}^r \right)^T$, whereas it refers to $\widehat {\mathbf{A}}^2$ for the estimation of $\mathbf{H}^s$. We also note that the means and variances of each column of the unknown channel matrices, needed in Algorithm~\ref{alg:vamp}, can be obtained from the unfolded forms of \eqref{noise_term}. Then, the unknown channels can be reconstructed from the means of the estimated channel vectors. The prior means $\widehat{\mathbf{h}}_{(0),c,{\rm mean}}$ and variances $\widehat{\mathbf{h}}_{(0),c,{\rm var}}$ of $\widehat{\mathbf{h}}_c$, $\forall$$c=1,2,\ldots,C$ are initialized using the random Gaussian distribution.
	\begin{algorithm}[!t]
		\caption{\textbf{VAMP-Based Computation of $\hat{\mathbf{H}}^r_{(i)}$ ($C=K$) or $\hat{\mathbf{H}}^s_{(i)}$ ($C=M$)}}
		\label{alg:vamp}
		\begin{algorithmic}[1]
			\Require Observation vectors $[\widetilde{\mathbf{Z}}]_{:,c}$ $\forall$$c\!=\!1,2,\!\ldots\!,C$, measurement matrix $\widehat {\mathbf{A}}_{(i-1)}, R \triangleq {\text{rank}}(\widehat {\mathbf{A}}_{(i-1)}), \sigma^2, \varepsilon>0$, and $I_{\rm max}$,  the prior means $\widehat{\mathbf{h}}_{(0),c,{\rm mean}}$ and the prior variances $\widehat{\mathbf{h}}_{(0),c,{\rm var}}$.
			\For {$c=1,2,\ldots,C$}
			\State \textbf{Initialization:} $\mathbf{r}_{(0),1}\!,\!\gamma_{(0),1}$, and iteration counter $p\!=\!0$.
			\State Compute the SVD $\widehat {\mathbf{A}}^1_{(i-1)} = \mathbf{US}{\mathbf{V}^H}$ and set ${\mathbf{\tilde y}}_{c} = \sigma^{-2}{\mathbf{S}^H}{\mathbf{U}^H}[\widetilde{\mathbf{Z}}]_{:,c}$.
			\Repeat
			\State Set ${\eta}_{(p)} = N^{-1}\sum_{j=1}^p 1 / \widehat{\mathbf{h}}_{(j),c,{\rm var}} + \gamma_{(p),1}$.
			\State Compute $\mathbf{\hat h}_{(p),1} \!\!\!\!= \!\eta_{(i)}^{\!-\!1}\!\left(\!\widehat{\mathbf{h}}_{(0),c,{\rm mean}}/\widehat{\mathbf{h}}_{(0),c,{\rm var}} \!+\! \mathbf{r}_{(p),1}\gamma_{(p),1}\right)$.
			\State Set $\gamma_{(p),2} = {\eta}_{(p)} - \gamma_{(p),1}$.
			\State Compute $\mathbf{r}_{(p),2} \!= \!\gamma_{(i),2}^{-1}\left(\mathbf{\hat h}_{(p),1}{\eta}_{(p)} - \mathbf{r}_{(p),1}\gamma_{(p),1}\right)$.
			\State Compute $\mathbf{D}_{(p)} = \left(\sigma^{-2}\mathbf{S}^H\mathbf{S} + \gamma_{(p),2}\mathbf{I}_N\right)^{-1}$.
			\State Compute $\mathbf{\hat h}_{(p),2} = \mathbf{V}\mathbf{D}_{(p)}\left(\mathbf{\tilde y}_{c} + \gamma_{(p),2}\mathbf{V}^H\mathbf{r}_{(i),2}\right)$.
			\State Set $\alpha_{(p)} \!=\! \gamma_{(p),2}R^{-1}\sum_{r=1}^N \left(\sigma^{-2}[\mathbf{S}]_{r,r}^2+\gamma_{(p),2}\right)^{-1}$.
			\State Compute $\mathbf{r}_{(p+1),1} \!\!\! = \!\left(\!\mathbf{\hat h}_{(p),2} \!-\! \alpha_{(p)}\mathbf{r}_{(p),2}\!\right)\!\!\left(\!1\!-\!\alpha_{(p)}\!\right)^{\!-\!1}$.
			\State Set $\gamma_{(p+1),1} = \gamma_{(p),2}\left(1-\alpha_{(p)}\right)\alpha_{(p)}^{-1}$.
			\State Set $p = p + 1$. 
			\Until $p > I_{\rm max}$ or ${\left\| \mathbf{r}_{(p+1),1} - \mathbf{r}_{(p),1}\right\|^2} < \kappa {\left\| \mathbf{r}_{(p),1} \right\|^2}$.
			\EndFor
	\Ensure The posterior means and variances of $\widehat{\mathbf{h}}_{c}$ $\forall$$c=1,2,\ldots,C$.
		\end{algorithmic}
	\end{algorithm}

The algorithmic steps of Algorithm~\ref{alg:vamp} including VAMP's transition variables can be divided into two parts. The denoising part includes Steps from $4$ to $7$, while the MMSE estimation part encompasses Steps $8$-$12$. In Step $2$ of each algorithmic iteration,  $\mathbf{r}_{(0),1}$ and $\gamma_{(0),1}$ are initialized using the MMSE estimation. Algorithm~2 outputs the posterior means and variances of $\widehat{\mathbf{h}}_c$ $\forall$$c=1,2,\ldots,C$, from which the means are used to construct $\hat{\mathbf{H}}^r_{(i)}$ when $C=K$ and $\hat{\mathbf{H}}^s_{(i)}$ for $C=M$. The proposed VAMP-based CE is summarized in Algorithm~3.
\begin{algorithm}[!t]\label{alg:als}\caption{\textbf{Proposed Iterative CE Based on VAMP}}
\label{alg:vamp_ce}
\begin{algorithmic}[1]
  \Require A feasible $\mathbf{\Phi}$, $\kappa>0$, and the number of maximum algorithmic iterations $I_{\rm max}$.
	\State \textbf{Initialization:} $\widehat{\mathbf{H}}^s_{(0)}$ is obtained from the $N$ non-zero eigenvalues of $(\mathbf{Z}^{2})^H \mathbf{Z}^2$, and set the algorithmic iteration as $i=1$.
	\For {$i=1,2,\ldots,I_{\rm max}$}
	\State Construct the channel $\widehat{\mathbf{H}}^r_{(i)}$ by Algorithm~\ref{alg:vamp}.
	\State Construct the channel $\widehat{\mathbf{H}}^s_{(i)}$ by Algorithm~\ref{alg:vamp}.
	\State \textbf{Until} {$\|\widehat{\mathbf{H}}^{r}_{(i)}-\widehat{\mathbf{H}}^{r}_{(i-1)} \|_{F}^{2}\|\widehat{\mathbf{H}}^{r}_{(i)} \|_{F}^{-2} \leq \kappa$ and $\|\widehat{\mathbf{H}}^{s}_{(i)}-\widehat{\mathbf{H}}^{s}_{(i-1)} \|_{F}^{2}\|\widehat{\mathbf{H}}^{s}_{(i)} \|_{F}^{-2} \leq \kappa$} or $i>I_{\rm max}$.
  \EndFor
\Ensure $\widehat{\mathbf{H}}^r_{(i)}$ and $\widehat{\mathbf{H}}^s_{(i)}$ that are the estimations of $\mathbf{H}^r$ and $\mathbf{H}^s$, respectively.
\end{algorithmic}
\end{algorithm}

\subsection{Feasibility Conditions}
To ensure that both proposed algorithms, which are based on the PARAFAC decomposition, yield a solution, some of the system parameters need to meet necessary and sufficient conditions. Based on the system identifiability conditions of \cite{rong2012channel}, the following lemma provides the feasibility conditions for both designed CE algorithms.
\begin{lemma} \label{lemma1}
The following system settings need to be met in order to mitigate the inherit ambiguity with the PARAFAC decomposition in \eqref{scalar_decomp}:
	\begin{equation}\label{eq:feasibility}
	  \operatorname{min}\left(M, K\right) \geq N.
	\end{equation}
\end{lemma}
This means that it needs to hold $M,K \geq N$ for the number of antennas, mobile users, and RIS unit elements, as well as the number of training RIS phase configurations $P$ should be less or equal to $N$. For those cases, the triple $(\mathbf{\Phi}, \mathbf{H}^s, \mathbf{H}^r)$ are unique up to some permutation and scaling ambiguities.
	
\textit{Proof:} According to the identifiability theorem of the PARAFAC decomposition, if it holds for $\mathbf{\Phi}$, $\mathbf{H}^s$, and $\mathbf{H}^r$ that:
	\begin{equation} \label{equ:iden}
	k_{\mathbf{\Phi}}+k_{\mathbf{H}^s}+k_{\mathbf{H}^r}\geq 2N+2,
	\end{equation}
then, the triple $(\mathbf{\Phi}, \mathbf{H}^s, \mathbf{H}^r)$ is unique up to permutation and scaling ambiguities, i.e.:
	\begin{equation}
	\widehat{\mathbf{H}}^r\!=\!\mathbf{H}^r \mathbf{\Pi} \mathbf{\Delta}_{1},   \widehat{\mathbf{\Phi}}\!=\!\mathbf{\Phi}\mathbf{\Pi} \mathbf{\Delta}_{2},    (\widehat{\mathbf{H}}^{s})^T\!=\!\left(\mathbf{H}^{s}\right)^T \mathbf{\Pi} \mathbf{\Delta}_{3},
	\end{equation}
where $\mathbf{\Pi}$ is an $N \times N$ permutation matrix and $\mathbf{\Delta}_{i}$, with $i=1,2$, and $3$, are $N \times N$ diagonal (complex) scaling matrices such that $\mathbf{\Delta}_{1} \mathbf{\Delta}_{2} \mathbf{\Delta}_{3}=\mathbf{I}_{N}$. According to the considered channel model, $\mathbf{H}^r$ and $\mathbf{H}^s$ are full rank and we choose $\mathbf{\Pi}$ to have the highest $k$-rank, hence \eqref{equ:iden} deduces to
\begin{equation}
	\min(P,M)+\min(N,M)+\min(K,N) \geq 2N+2.
\end{equation}
Using the latter expression and \cite[Theorem~1]{rong2012channel} completes the proof.
	
\textit{Remark:} It is evident from Lemma~1 that, for using the proposed ALS- and VAMP-based CE algorithms, the number of RIS unit elements $N$ cannot be greater than $M$ or $K$. However, it is expected in practice, that RISs will have large $N$, and particularly, larger than the number $K$ of BS antenna elements and/or the number $M$ of mobile users. In such cases, the feasibility conditions of Lemma~1 are not met. To address this conflict, both proposed algorithms need to be deployed in distinct portions of the RIS in the following way. The $N$-element RIS should be first partitioned in groups of non-overlapping sub-cells, such that each group's number of elements meets the feasibility conditions (i.e., less or equal to $M$ and $K$) \cite{9039554}. Then, both proposed algorithms can be used for each group to estimate portions of the intended channels, that can be finally combined to structure the desired CE. For example, an RIS with $64$ elements in total, a BS with $K=8$ antennas and $M=8$ mobile users, where $N \gg (M,K)$ that does not meet the condition of Lemma 1. However, the RIS can be split into $8$ non-overlapping sub-RISs each having $8$ elements to meet Lemma 1, and the proposed algorithms can be applied for estimating the channels referring to each sub-RIS. The proper concatenation of the latter estimations will yield the desired CE for the whole $64$-element RIS.

\subsection{Computational Complexity}

\begin{table}
	\caption{Computational Complexity of the Two Proposed CE Algorithms.}
	\centering
	\scalebox{0.8}{
	\begin{tabular}{|c|c|c|}
		\hline  &ALS&VAMP\\
		\hline  $\hat{\mathbf{H}}^s$&$\mathcal{O}\left(N^3+4N^2KP-NKP \right)$&$\mathcal{O}\left(M(5N^2-N)\right)$ \\
		\hline  $\hat{\mathbf{H}}^r$&$\mathcal{O}\left(N^3+4N^2MP-NMP\right)$ &$\mathcal{O}\left(K(5N^2-N)\right)$\\
		\hline
		$\textrm{Total}$ & $\mathcal{O}\left(2N^3+4N^2P(M+K)-NP(M+K)\right)$ & $\mathcal{O}((M+K)(5N^2-N))$\\
		\hline
	\end{tabular}}
	\label{table:complexity}
\end{table}

In Table~\ref{table:complexity}, we summarize the per-iteration computational complexity of two proposed algorithms. The complexity of  ALS-based algorithm is dominated by involved matrix inverse computations: the computational complexities of $\mathbf{H}^s$ and $\mathbf{H}^r$ are $\mathcal{O}\left(N^3+4N^2KP-NKP \right)$ and $\mathcal{O}\left(N^3+4N^2MP-NMP\right)$, respectively, as shown in Steps $3$ and $4$ of Algorithm~1.  Thus, the total computational complexity of the ALS-based CE is $\mathcal{O}\left(2N^3+4N^2P(M+K)-NP(M+K)\right)$. The VAMP-based algorithm is mainly dominated by matrix-vector multiplications in Step $10$ in Algorithm~2 \cite{8713501}. Thus,  the complexities of computing $\mathbf{H}^s$ and $\mathbf{H}^r$ are $\mathcal{O}\left(M(5N^2-N)\right)$ and $\mathcal{O}\left(K(5N^2-N)\right)$, respectively, in each inner iteration of Algorithm~2. Similarly, the total complexity of this technique can be approximated as $\mathcal{O}((M+K)(5N^2-N))$. Clearly, the VAMP algorithm has the less computational complexity than ALS due to the absence of the matrix inverse computation, and shows its superiority especially for larger matrices.

	\section{Cram\'er-Rao bound analysis}\label{sec:crb}		
In this section, we derive the CRBs for the estimations for $\mathbf{H}^{s}$ and $\mathbf{H}^{r}$, as obtained using the proposed ALS algorithm. It is noted that ALS yields the maximum likelihood estimate for the considered zero-mean AWGN received signal model, which is known to be asymptotically unbiased \cite{942635}. This means that CRBs provide the theoretically optimal performance for the ALS CE algorithm. The proposed VAMP CE technique involves indeterministic iterations (see Algorithm 2), and hence, it is very difficult, if not impossible, to obtain its CRBs. We next present the CRB analysis for the ALS technique.
	
Recall the following mode-$1$, mode-$2$, and mode-$3$ forms of $\widetilde{\mathbf{Z}}$:
\begin{align}
	\widetilde{\mathbf{Z}}^1 &=(\left(\mathbf{H}^{s}\right)^{\mathrm T}\circ\mathbf{\Phi})\left(\mathbf{H}^{r}\right)^{T} + \widetilde{\mathbf{W}}^1,
	\\
	\widetilde{\mathbf{Z}}^2 &=(\mathbf{\Phi}\circ\mathbf{H}^r)\mathbf{H}^s + \widetilde{\mathbf{W}}^2,
  \\
	\widetilde{\mathbf{Z}}^3 &=(\mathbf{H}^r\circ\left(\mathbf{H}^{s}\right)^{\mathrm T})\mathbf{\Phi}^{\mathrm T} + \widetilde{\mathbf{W}}^3,
\end{align}
where the AWGN matrices $\widetilde{\mathbf{W}}^2\in\mathbb{C}^{KP \times M}$ and $\widetilde{\mathbf{W}}^3\in\mathbb{C}^{MK \times P}$ are obtained from $\widetilde{\mathbf{W}}$. We make use of the notations $\mathbf{a}_k\triangleq[(\mathbf{H}^{r})^T]_{:,k}$ and $\mathbf{b}_m\triangleq\left[\mathbf{H}^{s}\right]_{:,m}$ to express the likelihood functions of $\widetilde{\mathbf{Z}}$ in following three equivalent ways:
	\begin{equation}
	\begin{aligned}
	L(\widetilde{\mathbf{Z}}) \! &= \! \frac{1}{({\pi \sigma^{2}} )^{K M P}} \! \exp \! \left\{ \! -\frac{1}{ \sigma^{2}} \! \sum_{k=1}^{K} \! \left\|[\widetilde{\mathbf{Z}}^1]_{:,k} \!-\! ( \!(\mathbf{H}^{s})^{\mathrm T} \! \odot \! \mathbf{\Phi} \!) \mathbf{a}_k \! \right\|^{2}\right\} \\
	\! &= \! \frac{1}{({\pi \sigma^{2}} )^{K M P}} \! \exp \! \left\{ \! -\frac{1}{ \sigma^{2}} \! \sum_{m=1}^{M}\left\|[\widetilde{\mathbf{Z}}^2]_{:,m}\!-\!(\!\mathbf{\Phi} \! \odot \! \mathbf{H}^{r} \!) \mathbf{b}_m \! \right\|^{2}\right\} \\
	\! &=\! \frac{1}{\!(\!{\pi \! \sigma^{2}} \!)\!^{K \!M \!P}} \! \exp \! \left\{ \! - \!\frac{1}{ \sigma^{2}} \! \sum_{p=1}^{P} \! \left\| \![ \widetilde{\mathbf{Z}}^3]_{:,p}\!-\!(\!\mathbf{H}^{r} \! \odot \! (\mathbf{H}^{s} \!)^{\!\mathrm T} \!) \! \left[ \!\mathbf{\Phi}^T\!\right]_{:,p} \! \right\|^{\!2} \!\right\}.
	\end{aligned}
	\end{equation}
To derive a meaningful CRB for the proposed ALS estimator by removing the scaling ambiguity, we fix $\mathbf{H}^s$ such that $[\mathbf{H}^s]_{:,1}=\mathbf{1}_N$. In this way, the number of unknown complex parameters in this matrix is $(K+M-1)N$ instead of $(K+M)N$. To derive the CRB of complex parameters, we also introduce the following complex parameter vector \cite{van1994cramer} 
	\begin{equation}
	\boldsymbol{\theta}\!\triangleq\!\left[\!\mathbf{a}_{1}^{\mathrm{T}}, \!\ldots\!, \mathbf{a}_{K}^{\mathrm{T}}, \mathbf{b}_{2}^{\mathrm{T}}, \!\ldots\!, \mathbf{b}_{M}^{\mathrm{T}},  \mathbf{a}_{1}^{H}, \!\ldots\!, \mathbf{a}_{K}^{H}, \mathbf{b}_{2}^{H}, \!\ldots\!, \mathbf{b}_{M}^{H} \!\right].
	\end{equation}

The log-likelihood function of the unknowns in $\boldsymbol{\theta}$ can be expressed as
	\begin{equation}
	\begin{aligned}
	f(\boldsymbol{\theta}) &=\!-\!{{K M P} \ln ({\pi \sigma^{2}} )} \! - \! \frac{1}{ \sigma^{2}} \sum_{k=1}^{K}\left\|[\mathbf{Z}_{1}]_{:,k}\! - \!((\mathbf{H}^{s})^{\mathrm T} \odot \mathbf{\Phi}) \mathbf{a}_k \right\|^{2}  \\
	&=\!- \!{{K M P} \ln ({\pi \sigma^{2}} )} \! - \!\frac{1}{ \sigma^{2}} \sum_{m=1}^{M}\left\|[\mathbf{Z}_{2}]_{:,m} \!- \!(\mathbf{\Phi} \odot \mathbf{H}^{r}) \mathbf{b}_m \right\|^{2},
	\end{aligned}
	\end{equation}
and the complex Fisher Information Matrix (FIM) of $\boldsymbol{\theta}$ is given by
	\begin{equation} \label{equ:fim}
	\Omega(\boldsymbol{\theta})=\mathrm{E}\left\{\left(\frac{\partial f(\boldsymbol{\theta})}{\partial \boldsymbol{\theta}}\right)^{H}\left(\frac{\partial f(\boldsymbol{\theta})}{\partial \boldsymbol{\theta}}\right)\right\}.
	\end{equation}
Taking the partial derivatives of $f(\boldsymbol{\theta})$ with respect to the unknown parameters in $\boldsymbol{\theta}$ yields	
	\begin{equation}\label{equ:deri}
		\begin{aligned}
		&{\frac{\partial f(\boldsymbol{\theta})}{\partial [\mathbf{a}_{k}]_n}=\frac{1}{\sigma^{2}}\left([\mathbf{Z}_{1}]_{:,k}-((\mathbf{H}^{s})^T \odot \mathbf{\Phi}) \mathbf{a}_{k}\right)^{H} ((\mathbf{H}^{s})^{\mathrm T} \odot \mathbf{\Phi}) \mathbf{e}_{n}  },\\
		& {\frac{\partial f(\boldsymbol{\theta})}{\partial [\mathbf{b}_{m}]_n}=\frac{1}{\sigma^{2}}\left([\mathbf{Z}_{2}]_{:,m}-(\mathbf{\Phi} \odot \mathbf{H}^{r}) \mathbf{b}_{m}\right)^{H} (\mathbf{\Phi} \odot \mathbf{H}^{r}) \mathbf{e}_{n}  },\\
		& \frac{\partial f(\boldsymbol{\theta})}{\partial [\mathbf{a}_{k}]_n^{*}}=\left(\frac{\partial f(\boldsymbol{\theta})}{\partial a_{k, n}}\right)^{*},  \frac{\partial f(\boldsymbol{\theta})}{\partial [\mathbf{b}_{m}]_n^{*}}=\left(\frac{\partial f(\boldsymbol{\theta})}{\partial b_{m, n}}\right)^{*}.
		\end{aligned}
	\end{equation}
     Based on \eqref{equ:fim} and \eqref{equ:deri}, $E \{ \left([\mathbf{Z}_{1}]_{:,k} \!- \!((\mathbf{H}^{s})^{\mathrm T} \!\odot\! \mathbf{\Phi}) \mathbf{a}_{k}\right) \left([\mathbf{Z}_{1}]_{:,k}\!-\!((\mathbf{H}^{s})^T \!\odot\! \mathbf{\Phi}) \mathbf{a}_{k}\right)^{H} \}=\mathbf{I}$, and  $E \{\! \left([\mathbf{Z}_{2}]_{:,m}\!-\!(\mathbf{\Phi}\! \odot \!\mathbf{H}^{r}) \mathbf{b}_{m}\right)  \!\left([\mathbf{Z}_{2}]_{:,m}\!-\!(\mathbf{\Phi} \!\odot \!\mathbf{H}^{r}) \mathbf{b}_{m}\right) ^{H} \! \}\!=\! \mathbf{I}$. Due to that $\mathbf{e}_{n1}$ is independent of $\mathbf{e}_{n2}$ for $n1 \neq n2$, we have $E\left\{\frac{\partial f(\boldsymbol{\theta})}{\partial [\mathbf{a}_{k}]_{n_{1}}^{*}} \frac{\partial f(\boldsymbol{\theta})}{\partial [\mathbf{a}_{k}]_{n_{2}}}\right\} = \mathbf{0}$.  
  
  Hence, the FIM can be re-expressed in the compact form:
	\begin{equation}
	\mathbf{\Omega}(\boldsymbol{\theta})=\left[\begin{array}{cc}
	{\mathbf{\Psi}} & \mathbf{0} \\
	\mathbf{0} & {\mathbf{\Psi}^{*}}
	\end{array}\right],
	\end{equation}
where the all-zero matrix $\mathbf{0}$ and $\mathbf{\Psi}$ are of size $((K+M-1)N)\times((K+M-1)N)$. $\mathbf{\Psi}$'s elements are given by combining \eqref{equ:fim} and the following expressions:
	\begin{equation} \label{equ:crb_exp}
		\begin{aligned}
		&E\left\{\frac{\partial f(\boldsymbol{\theta})}{\partial [\mathbf{a}_{k_{1}}]_{n_{1}}^{*}} \frac{\partial f(\boldsymbol{\theta})}{\partial [\mathbf{a}_{k_{2}}]_{n_{2}}}\right\} \\  &=\frac{1}{\sigma^{2}} \mathbf{e}_{n_{1}}^{H}(\mathbf{H}^{s^\mathrm{T}} \odot \mathbf{\Phi})^{H}(\mathbf{H}^{s^\mathrm{T}} \odot \mathbf{\Phi}) \mathbf{e}_{n_{2}} \delta_{k_{1}, k_{2}}, \\
		&E\left\{\frac{\partial f(\boldsymbol{\theta})}{\partial [\mathbf{b}_{m_{1}}]_{n_{1}}^{*}} \frac{\partial f(\boldsymbol{\theta})}{\partial [\mathbf{b}_{m_{1}}]_{n_{1}}}\right\}\\
		& =\frac{1}{\sigma^{2}} \mathbf{e}_{n_{1}}^{H}(\mathbf{\Phi} \odot \mathbf{H}^{r})^{H}(\mathbf{\Phi} \odot \mathbf{H}^{r}) \mathbf{e}_{n_{2}} \delta_{m_{1}, m_{2}}, \\
		&  E\left\{\frac{\partial f(\boldsymbol{\theta})}{\partial [\mathbf{a}_{k}]_{n_{1}}^{*}} \frac{\partial f(\boldsymbol{\theta})}{\partial [\mathbf{b}_{m}]_{n_{2}}}\right\}\\
		& = \frac{1}{\sigma^{2}} \mathbf{e}_{n_{1}}^{H}(\mathbf{H}^{s^\mathrm{T}} \odot \mathbf{\Phi})^{H} E\left\{[\widetilde{\mathbf{W}}^{1}]_{:,k}  [\widetilde{\mathbf{W}}^{2}]_{:,m}^{H}\right\}  (\mathbf{\Phi} \odot \mathbf{H}^{r}) \mathbf{e}_{n_{2}},
		\end{aligned}
	\end{equation}
for $k_{1},k_{2}=1,2,\ldots,K$ and $n_{1},n_{2}=1,2,\ldots,N$. In the latter expression, $E\left\{[\widetilde{\mathbf{W}}^{1}]_{:,k}  [\widetilde{\mathbf{W}}^{2}]_{:,m}^{H}\right\}$ represents AWGN's covariance matrix, which is given by
	\begin{equation}\label{noise_matrix}
		\begin{aligned}
	&E\left\{[\widetilde{\mathbf{W}}^{1}]_{:,k}  [\widetilde{\mathbf{W}}^{2}]_{:,m}^{H}\right\}\\&=\! \sigma^{2} \!\! \! \!
	\begin{array}{c}
	\left[\! \begin{array}{ccccccccc}
	\! 0 & \!\! \cdots\! \! & \! 0 \! & \! \cdots\!  & \! 0\!  & \! \cdots \! & \! 0 \! &\!  \cdots\!  & \! 0\!  \\
	& & & \! \ddots\!  & & & & \\
	\! 0 & \!\! \cdots\! \! & \! 1\!  & \! \cdots\!  & \! 0\!  &\!  \cdots\!  &\!  0\!  & \! \cdots \! & \! 0\!  \\
	\! 0 \! & \! \cdots \! &\!  0 \! &\!  \cdots \! &\!  1 \! & \! \cdots \! &\!  0\!  & \! \cdots \! &\!  0 \! \\
	& & & \! \!\ddots\!\!  & & & & & \\
	\! 0 \!\! & \! \cdots\!  & \! 0\!  & \! \cdots\!  &\!  0\!  & \! \cdots\!  & \! 1\!  & \! \cdots & \! 0\!  \\
	& & & \!\! \ddots\!\!  & & & & & \\
	\! 0 \! \!& \! \!\cdots\!\!  &\!  0\!  & \! \cdots\!  & \! 0 \! & \! \cdots\!  & \! 0\!  & \! \cdots\!  &\!  0\! 
	\end{array} \! \right] \! 
	\begin{array}{c}
	\!\! \leftarrow {\!(m-1)P+1} \\
	\!\!\leftarrow {\!(m-1)P+p} \\
	\!\!\cdots \\
	\!\!\leftarrow {\!(m-1)P+P}
	\end{array} \\
	\begin{array}{cccccccc}
	\uparrow & & \uparrow &  & \uparrow & & &   \\
	k & & K+k & \cdots & (P-1)K+k & & &
	\end{array}
	\end{array}
        \end{aligned}
	\end{equation}
	
The CRB for the proposed unbiased ALS estimator for the unknown channels in $\boldsymbol{\theta}$ using the considered trilinear model in Section~II is obtained as   	
	\begin{equation}
	\mathbf{\Omega}^{-1}(\boldsymbol{\theta})=\left[\begin{array}{cc}
	{\mathbf{\Psi}^{-1}} & {\mathbf{0}}\\
	{\mathbf{0}} & {\left(\mathbf{\Psi}^{-1}\right)^{*}}
	\end{array}\right],
	\end{equation}
	where the inverse of $\mathbf{\Psi}$ is given by
	\begin{equation} \label{equ:crb_channels}
	\mathbf{\Psi}^{-1} = \left[\begin{array}{cc}
	\boldsymbol{\mathrm{CRB}}_{\mathbf{H}^r} & \mathbf{K}\\
	\mathbf{K}^H & \boldsymbol{\mathrm{CRB}}_{\mathbf{H}^s}
	\end{array}\right]
	\end{equation}
where $\boldsymbol{\mathrm{CRB}}_{\mathbf{H}^r}\in \mathbb{C}^{K N \times KN}$ is the CRB for the CE of $\mathbf{H}^r$, $\boldsymbol{\mathrm{CRB}}_{\mathbf{H}^s}\in \mathbb{C}^{((M-1) N) \times ((M-1)N)}$ is the CRB for the CE of $\mathbf{H}^s$, and $\mathbf{K} \in \mathbb{C}^{K N \times ((M-1) N)}$ represents the remaining submatrices. By using the formula for the inverse of a partitioned Hermitian matrix \cite{942635}, the targeted CRBs are derived as
	\begin{equation} \label{equ:crb_hr}
	\boldsymbol{\mathrm{CRB}}_{\mathbf{H}^r}=\left(\mathbf{\Psi}_{1}-\mathbf{\Psi}_{2} \mathbf{\Psi}_{3}^{-1} \mathbf{\Psi}_{2}^{H}\right)^{-1},
	\end{equation}
	\begin{equation} \label{equ:crb_hs}
	\boldsymbol{\mathrm{CRB}}_{\mathbf{H}^s}=\left(\mathbf{\Psi}_{3}-\mathbf{\Psi}_{2}^{H} \mathbf{\Psi}_{1}^{-1} \mathbf{\Psi}_{2}\right)^{-1},
	\end{equation}
	where ${\mathbf{\Psi}_{1}\in \mathbb{C}^{K N \times KN}}$, ${\mathbf{\Psi}_{3}\in \mathbb{C}^{(M-1) N \times (M-1)N}}$ and ${\mathbf{\Psi}_{2}\in \mathbb{C}^{K N \times (M-1)N}}$ partition matrix $\mathbf{\Psi}$ as
	\begin{equation}
	\mathbf{\Psi}=\left[\begin{array}{cc}
	{\mathbf{\Psi}_{1}} & {\mathbf{\Psi}_{2}} \\
	{\mathbf{\Psi}_{2}^{H}} & {\mathbf{\Psi}_{3}}
	\end{array}\right].
	\end{equation}

	
\section{Sum Rate Performance Computation}\label{sec:sum_rate}
In this section, we present the downlink sum rate performance formulas using the MRT, ZF, and MMSE precoding schemes that will be evaluated in the results' section later on.   The downlink channels can be obtained through the reciprocity of uplink channels, which are estimated using the proposed algorithms.   Considering flat fading channel conditions and assuming perfect channel availability for computing the precoding vectors $\mathbf{g}_{k}\in \mathbb{C}^{M \times 1}$, $\forall$$k=1,2,\ldots,K$ and the RIS phase configuration matrix $\mathbf{\Phi}$, the baseband received signals at the $K$ mobile users, when included in $\mathbf{y}\in\mathbb{C}^{K\times1}$, are given by
	\begin{equation}\label{equ:input_output}
	\mathbf{y} \triangleq\mathbf{H}^{r}  \mathbf{\Phi}   \mathbf{H}^{s} \mathbf{x} +\mathbf{w},
	\end{equation}
where $\mathbf{x} \triangleq \sqrt{p}_{u} \sum_{k=1}^{K}  \mathbf{g}_{k} s_{k}\in\mathbb{C}^{M\times1}$ denotes the transmitted signal with power $p_u$. This signal consists of the unit-power complex-valued information symbol $s_k$ (chosen from a discrete constellation set) for each $k$-th user, which is beamformed via the precoding vector $\mathbf{g}_{k}$. Vector $\mathbf{w}\in\mathbb{C}^{K\times1}$ in \eqref{equ:input_output} represents AWGN such that $\mathbf{w}\sim\mathcal{CN}(\mathbf{0}, \sigma^2 \mathbf{I}_K)$. By using the notations $\mathbf{H}\triangleq\mathbf{H}^{r}  \mathbf{\Phi}  \mathbf{H}^{s} \in \mathbb{C}^{K \times M}$ and $\mathbf{h}_{k}\triangleq[\mathbf{H}^{H}]_{:,k}$, the received signal at the $k$-th user (i.e., the $k$-th element of $\mathbf{y}$) can be expressed as
\vspace{-2mm}
    \begin{equation}
    \begin{aligned}
     y_k = \sqrt{p}_{u} \mathbf{h}_{k}^{H}  \mathbf{g}_{k} s_{k}  + \sqrt{p}_{u} \sum_{i \neq k}^{K} {\mathbf{h}_{k}^{H}   \mathbf{g}_{i} s_{i}}  +w_k,
    \end{aligned}
    \end{equation}
where $w_k$ is the $k$-th element of $\mathbf{w}$ in \eqref{equ:input_output}. It is noted that last two terms represent the interference-plus-noise term, which is a random variable having zero mean and variance $p_u \sum_{i \neq k}^{K} {| \mathbf{h}_{k}^{H}   \mathbf{g}_{i} |}^2 + \sigma^2$. We next model this term as additive Gaussian noise that is independent of $s_k$. 

For the case where either of the proposed PARAFAC-based CE techniques is deployed, the CEs $\widehat{\mathbf{H}}^{r}$ and $\widehat{\mathbf{H}}^{s}$ are used in \eqref{equ:input_output}. Obviously, when there exist CE errors, the achievable sum rate performance will be impacted. By defining the estimation error matrices for $\mathbf{H}^{r}$ and $\mathbf{H}^{s}$ as $\boldsymbol{\mathcal{E}}^{r}\in\mathcal{C}^{K\times N}$ and $\boldsymbol{\mathcal{E}}^{s}\in\mathcal{C}^{N\times M}$, respectively, the actual matrix $\mathbf{H}$ is given by
    \begin{equation}
    \begin{aligned}
    \mathbf{H}&=\left ( \widehat{\mathbf{H}}^{r}-\boldsymbol{\mathcal{E}}^{r} \right )  \mathbf{\Phi}  \left (  \widehat{\mathbf{H}}^{s}-\boldsymbol{\mathcal{E}}^{s} \right ) =  \widehat{\mathbf{H}}^{r} \mathbf{\Phi}  \widehat{\mathbf{H}}^{s} - \boldsymbol{\mathcal{E}},
    \end{aligned}
    \end{equation}
where $\boldsymbol{\mathcal{E}} \triangleq \boldsymbol{\mathcal{E}}^{r} \mathbf{\Phi}  \widehat{\mathbf{H}}^{s} +  \widehat{\mathbf{H}}^{r} \mathbf{\Phi}  \mathbf{\mathcal{E}}^{s}  -  \boldsymbol{\mathcal{E}}^{r}  \mathbf{\Phi} \boldsymbol{\mathcal{E}}^{s}$. Using the latter expressions, the received signal model with the estimated channels is derived as
    \begin{equation}
    \begin{aligned} \label{eq39}
    y_k \!=\!\!\sqrt{p}_{u} \!\widehat{\mathbf{h}}_{k}^{H} \! \widehat{\mathbf{g}}_{k}\! s_{k}\! \!+ \!\!\sqrt{p}_{u} \!\sum_{i \neq k}^{K} \!{\widehat{\mathbf{h}}_{k}^{H}  \! \widehat{\mathbf{g}}_{i} s_{i}}  \!\!- \!\!\sqrt{p}_{u} \!\sum_{i =1}^{K}\! {\widehat{\mathbf{h}}_{k}^{H} \!  {\boldsymbol{\varepsilon}}_{i}\! s_{i}}  \!  \! +\! \! w_k,
    \end{aligned}
    \end{equation}
  where $\widehat{\mathbf{g}}_{k}\in \mathbb{C}^{M \times 1}$ $\forall$, $k=1,2,\ldots,K$ are the BS precoding vectors that are now obtained based on the estimated channels, and  $\boldsymbol{\varepsilon}_i$ is $\boldsymbol{\mathcal{E}}$'s $i$-th column. This formula includes the intended signal in the first term and the remaining terms are treated as interference-plus-noise. We henceforth define the power of the third term in \eqref{eq39} as $\epsilon\triangleq{p}_{u} \sum_{i =1}^{K} \|{\widehat{\mathbf{h}}_{k}^H {\boldsymbol{\varepsilon}}_{i} }\|^2$. 

In the following, we assume that $\widehat{\mathbf{H}}^{s}$ and $\widehat{\mathbf{H}}^{r}$ has been first used for computing $\mathbf{\Phi}$ via the fixed point method proposed in \cite{8855810}. This method finds in an iterative way the phase matrix that matches the directions of the estimated channels, requiring moderate computational complexity. Then, the cascade channel $\widehat{\mathbf{H}}\triangleq\widehat{\mathbf{H}}^{r} \mathbf{\Phi}  \widehat{\mathbf{H}}^{s}\in \mathbb{C}^{K \times M}$ will be used to design the BS precoding vectors $\widehat{\mathbf{g}}_{k}$ $\forall$$k=1,2,\ldots,K$.
 
 
\subsection{The estimation of phase matrix}
For simplicity, we do not design a specific phase matrix $\mathbf{\Phi}$ for different precoding matrices. In contrast, we estimate a phase matrix with the aid of estimated channels. The phase matrix can be further optimized using the maximum ratio combining, which is equivalent to project the channel vectors on the phase matrix with the matched direction. 

The optimization problem can be formulated as 
\vspace{-2mm}
\begin{equation}
\begin{aligned}
&\max_{\mathbf{\Phi}}  \quad \sum_{k=1}^{K} \|{[\mathbf{h}_{k}^{r}]^H}\mathbf{\Phi}\mathbf{H}^{s} \|^2
\\
&\text {s.t.} \quad  |\phi_n|^2=1, n=1,2,\ldots,N
\end{aligned}
\label{equ:phi_ls}
\end{equation}
where ${\mathbf{h}^{r}_k}$ is the $k$-th column of $[\mathbf{H}^{r}]^H$.	Let $[\mathbf{h}_{k}^{r}]^H\mathbf{\Phi} \mathbf{H}^{s} =\mathbf{v}^H \text{diag}({\mathbf{h}_{k}^{r}}) \mathbf{H}^{s} =\mathbf{v}^H \mathbf{C}_k$, where $\mathbf{v}=\left [\phi_{1}, \ldots, \phi_{N}\right ]^{H}$, $\mathbf{C}_k=\operatorname{diag}({\mathbf{h}_{k}^{r}}) \mathbf{H}^{s}$. Thus, \eqref{equ:phi_ls} can be rewritten as
\vspace{-2mm}
\begin{equation}
\begin{aligned}
&\max_{\mathbf{v}} \quad {\sum_{k=1}^{K} \mathbf{v}^{H}\mathbf{C}_{k}^{\prime}} \mathbf{v}
\\
& \;\text{s.t.}\;  \quad |\mathbf{v}_n|^2=1, n=1,2,\ldots,N
\end{aligned}
\label{equ:v_ori}
\end{equation}
where ${\mathbf{C}_k^{\prime}} = {\mathbf{C}_k \mathbf{C}_k^H}$. Due to that ${\sum_{k=1}^{K} \mathbf{v}^{H}\mathbf{C}_{k}^{\prime}\mathbf{v}}={ \mathbf{v}^{H} \tilde{\mathbf{C}} \mathbf{v}}$, with $\tilde{\mathbf{C}} = \sum_{k=1}^{K} {\mathbf{C}_k^{\prime} }$.

Consequently, equation \eqref{equ:v_ori} can be further computed as
\vspace{-2mm}
\begin{equation}
\begin{aligned}
&\max_{\mathbf{v}} \quad \mathbf{v}^{H}\tilde{\mathbf{C}} \mathbf{v}
\\
& \;\text{s.t.}\;  \quad |\mathbf{v}_n|^2=1, n=1,2,\ldots,N
\end{aligned}
\label{equ:v_eq}
\end{equation}

Based on the fixed point iteration in \cite{8855810}, this optimization problem can be solved in an iterative way, which has the lower computational complexity, as shown in Algorithm~4.

\begin{algorithm}[htb]  
	\textbf{Algorithm 4 Iterative Phase Matrix Estimation  \cite{8855810}}
	\begin{enumerate}
		\item  Construct an initial $\widehat{\mathbf{v}}^{(0)}$ and set $t=0$
		\item  \textbf{repeat}
		\item  $\quad$ Perform the iteration according to $ \widehat{\mathbf{v}}^{ (t+1)} \!=\!\operatorname{unt}\!\left(\!\tilde{\mathbf{C}} \widehat{\mathbf{v}}^{(t)}\!\right)$
		\item  $t \leftarrow t+1$
		\item  \textbf{until} $\left\|\tilde{\mathbf{C}} \widehat{\mathbf{v}}^{(t+1)}\right\|_{1}-\left\|\tilde{\mathbf{C}} \widehat{\mathbf{v}}^{(t)}\right\|_{1} \leq \kappa$
		\item  Take the first $N$ elements of $\left(\widehat{\mathbf{v}}^{(t+1)}\right)^{*}$ as the main diagonal elements of matrix $\mathbf{\Phi}$
	\end{enumerate}
\end{algorithm}

where $\operatorname{unt}(\mathbf{a})$ forms a vector whose elements are $\frac{a_{1}}{\left|a_{1}\right|}, \cdots, \frac{a_{n}}{\left|a_{n}\right|}$.
With the solution $\widehat{\mathbf{v}}$, we obtain the phase matrix $\mathbf{\Phi}=\text{diag}(\widehat{\mathbf{v}})$, which can be further employed in deriving the sum rate with different precoding schemes. 
 

\subsection{MRT Precoding}
 When the BS applies MRT precoding, it holds $\widehat{\mathbf{g}}_k=\widehat{\mathbf{h}}_{k}\triangleq[\widehat{\mathbf{H}}^{H}]_{:,k}$, $\forall$$k=1,2,\ldots,K$.  In this case, the conceived by the system as the achievable rate for the $k$-th user is computed as
    \begin{equation} \label{sm_mrc_es}
    \begin{aligned}
    \widehat{\mathcal{R}}_{k}^{({\rm MRT})} \! = \! \mathbb{E} \left \{ \! \log _{2}\left(1 \!+ \!\frac{ p_u \left\|\widehat{\mathbf{h}}_{k}\right\|^{2}}{ p_u  \!\sum_{i \neq k}^{K}\!\left|\!\widehat{\mathbf{h}}_{k}^H  \widehat{\mathbf{h}}_{i}\!\right|^{2}\! +\!  \epsilon \!+\! \sigma^2}\right)  \! \right  \}.
    \end{aligned}
    \end{equation}
For the perfect CE case, the BS sets $\mathbf{g}_k= \mathbf{h}_k$ $\forall$$k=1,2,\ldots,K$ and the achievable downlink rate for the $k$-th user becomes:
    \begin{equation} \label{sm_mrc_pf}
    \begin{aligned}
    \mathcal{R}_{k}^{({\rm MRT})}=\mathbb{E} \left \{ \log _{2}\left(1+\frac{ p_u  \left\|\mathbf{h}_{k}\right\|^{2}}{ p_u \sum_{i \neq k}^{K}\left|\mathbf{h}_{k}^H  \mathbf{g}_{i}\right|^{2}+ \sigma^2}\right) \right \}.
    \end{aligned}
    \end{equation}
		
\subsection{ZF Precoding}
This precoding scheme intends at eliminating interference among different users by setting the precoding matrix as $\widehat{\mathbf{G}}=\widehat{\mathbf{H}}^{H}\left(\widehat{\mathbf{H}} \widehat{\mathbf{H}}^{H} \right)^{-1}$ for the case of CE. In this case, it holds $\widehat{\mathbf{h}}_k^H\widehat{\mathbf{g}}_i=\delta_{k,i}$. It is noted that the $k$-th column of the latter matrix is the precoding vector $\widehat{\mathbf{g}}_k$, which is applied to symbol $s_k$ intended for the $k$-th user. The conceived by the system as the achievable rate for the $k$-th user is given by
   \begin{equation} \label{sm_zf_es}
   \begin{aligned}
   \widehat{\mathcal{R}}_{k}^{({\rm ZF})}= \mathbb{E} \left \{ \log _{2}\left(1+\frac{ p_u }{\epsilon+ \sigma^2}\right)  \right \},
   \end{aligned}
   \end{equation}
whereas for the perfect channel knowledge case, the sum rate for $k$-th user is
    \begin{equation}
    \mathcal{R}_{k}^{({\rm ZF})}=\log _{2}\left(1+\frac{ p_u }{ \sigma^2}\right).
    \end{equation}

\subsection{MMSE Precoding}
Using $\widehat{\mathbf{H}}$ at BS, the precoding matrix for the MMSE precoding case is designed as
\begin{equation}\label{equ:mmse_g}
    \widehat{\mathbf{G}}=\left\{\begin{array}{ll}
    \widehat{\mathbf{H}}^{H}\left(\widehat{\mathbf{H}} \widehat{\mathbf{H}}^{H}+\frac{\sigma^2}{p_u} \mathbf{I}_K\right)^{-1}, &  \text{for}\,\,\, K \leq M  \\
    \left(\widehat{\mathbf{H}}^{H} \widehat{\mathbf{H}}+\frac{\sigma^2}{p_u}\mathbf{I}_M\right)^{-1} \widehat{\mathbf{H}}^{H}, &   \text{for}\,\,\, K > M
    \end{array}\right..
\end{equation}
In this case, the achievable downlink rate for the $k$-th user is computed by the system as
	\begin{equation} \label{sm_mmse_es}
	\begin{aligned}
	\widehat{\mathcal{R}}_{k}^{({\rm MMSE})} \!= \! \mathbb{E} \left \{\! \!\log _{2}\!\!\left(\! \! 1\! \!+ \!\! \frac{ p_u \left|[\widehat{\mathbf{H}}\widehat{\mathbf{G}}]_{k,k}\right|^{2}}{ p_u  \!\sum_{i=1, i \neq k}^{K}\!\left|\!\widehat{\mathbf{h}}_{k}^H  \widehat{\mathbf{g}}_{i}\right|^{2} \!+ \! \epsilon \!+\! \sigma^2}\!\!\right)  \! \!\right \}\!,\!
	\end{aligned}
	\end{equation}
while for the perfect channel knowledge case, the sum rate for $k$-th user is given by
	\begin{equation} \label{sm_mmse_pf}
	\begin{aligned}
	\mathcal{R}_{k}^{({\rm MMSE})}\!=\!\mathbb{E} \! \left \{ \!\log _{2}\left(\! 1 \!+ \!\frac{ p_u \! \left| \![\mathbf{H}\mathbf{G}]_{k,k}\!\right|^{2}}{ p_u \sum_{i=1, i \neq k}^{K}\left|\mathbf{h}_{k}^H  \mathbf{g}_{i}\right|^{2} \!+\! \sigma^2}\!\right)\! \!\right \}\!.\!
	\end{aligned}
	\end{equation}
In the latter expression, $\mathbf{G}$ is derived as in \eqref{equ:mmse_g} by replacing $\widehat{\mathbf{H}}$ with $\mathbf{H}$.
	
 \section{Performance Evaluation Results}\label{sec:simulation}
In this section, we present computer simulation results for the NMSE performance of the proposed CE algorithms as well as their achievable downlink sum rates.

\subsection{NMSE of CE}
We have particularly simulated the NMSE of our channel estimator using the metrics $\|\mathbf{H}^{s}-\widehat{\mathbf{H}}^{s}\|^{2}\|\mathbf{H}^{s}\|^{-2}$ and $\|\mathbf{H}^{r}-\widehat{\mathbf{H}}^{r}\|^{2}\|\mathbf{H}^{r}\|^{-2}$. The scaling ambiguity of the proposed algorithms has been removed by normalizing the first column of the channel matrices. During the phase of estimating the channels $\mathbf{H}^{s}$ and $\mathbf{H}^{r}$, $\mathbf{\Phi}$ is designed as the discrete Fourier transform matrix \cite{rong2012channel}, which satisfies $\mathbf{\Phi}^H\mathbf{\Phi}=\mathbf{I}_{N}$. All NMSE curves were obtained after averaging over $2000$ independent Monte Carlo channel realizations. We have used $\kappa=10^{-5}$ and $I_\text{max}=20$ in all presented NMSE performance curves.

\begin{figure} \vspace{-1mm}
    	\begin{center}
    		\includegraphics[width=0.45\textwidth]{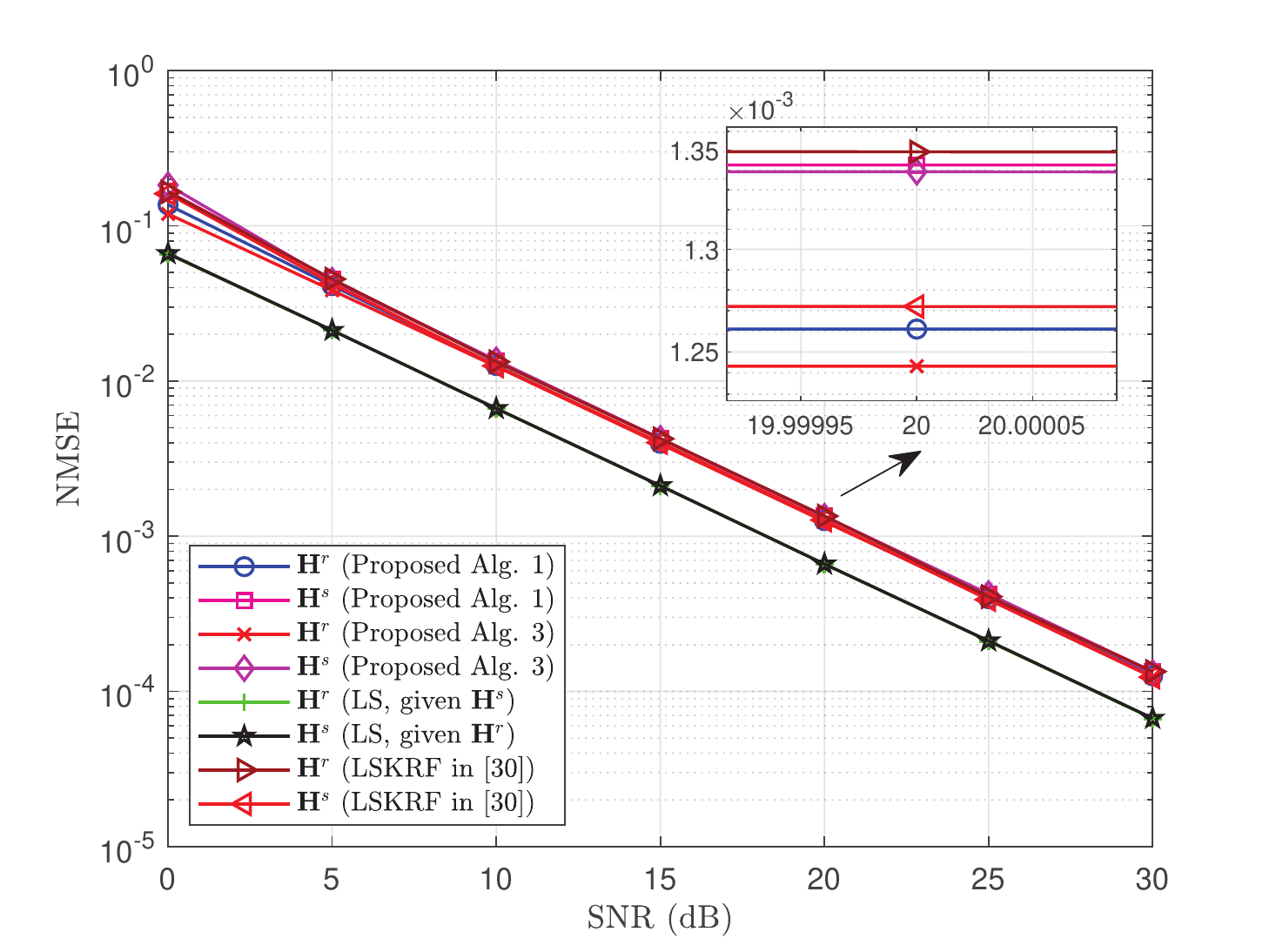}  \vspace{-2mm}
    		\caption{NMSE performance comparisons between the proposed algorithms, genie-aided LS estimation, and LSKRF \cite{9104260}  versus the SNRs in dB for $M=K=T=N=P=16$.}
    		\label{fig:com_ls_lskrf(all16)} \vspace{-6mm}
    	\end{center}
    \end{figure}

\begin{figure} \vspace{-1mm}
    	\begin{center}
    		\includegraphics[width=0.45\textwidth]{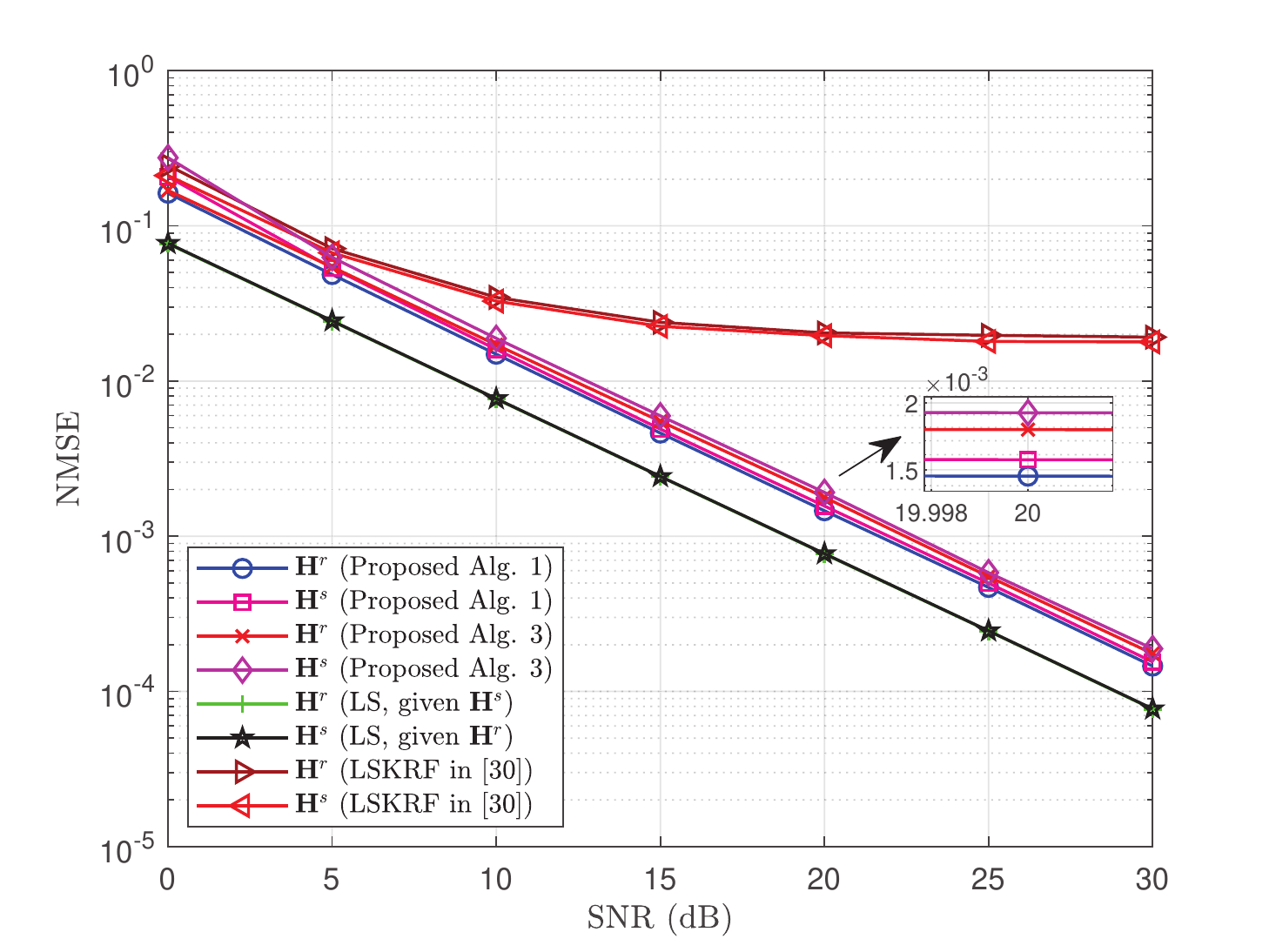}  \vspace{-2mm}
    		\caption{NMSE performance comparisons between the proposed algorithms, genie-aided LS estimation, and LSKRF \cite{9104260}  versus the SNR in dB for $M=K=T=N=16$, $P=14$.}
    		\label{fig:com_ls_lskrf_P14} \vspace{-6mm}
    	\end{center}
    \end{figure}

NMSE performance comparisons between the proposed ALS-based CE (Alg. 1), VAMP-based CE (Alg. 3), genie-aided LS estimation, and the LSKRF method \cite{9104260} are illustrated in Figs$.$~\ref{fig:com_ls_lskrf(all16)} and~\ref{fig:com_ls_lskrf_P14}. We have set $M=K=T=N=P=16$ in Fig$.$~\ref{fig:com_ls_lskrf(all16)}, while $M=K=T=N=16$ and $P=14$  in Fig$.$~\ref{fig:com_ls_lskrf_P14}. As shown in both figures, there is about $2.5$ dB performance gap between each of the proposed algorithms and the genie-aided LS estimation that estimates one unknown channel based on the genie-aided knowledge of another channel. In Fig$.$~\ref{fig:com_ls_lskrf(all16)}, the proposed algorithms exhibit similar NMSE performance with the LSKRF scheme, however, in Fig$.$~\ref{fig:com_ls_lskrf_P14}, they outperform LSKRF for around an order of magnitude lower NMSE at $\text{SNR}=20$ dB. Specifically, in Fig$.$~\ref{fig:com_ls_lskrf_P14}, the performance of LSKRF nearly converges at $\text{SNR}=20$ dB with $\text{NMSE}=0.02$. In contrast, the proposed algorithms achieve $\text{NMSE}=0.002$ at $\text{SNR}=20$ dB. Furthermore, while SNR increases, the gap between them increases. This happens because our proposed algorithms have less restrictive requirements than LSKRF, and are more robust. Even with less training phase matrices, the proposed algorithms can approximate the estimation performance of the genie-aided LS estimation. From these figures, it is obvious that both proposed algorithms share similar performance under the different examined settings. Thus, we use only our ALS CE in the following simulations for performance comparisons.

\begin{figure}\vspace{-1mm}
    	\begin{center}
    		\centerline{\includegraphics[width=0.45\textwidth]{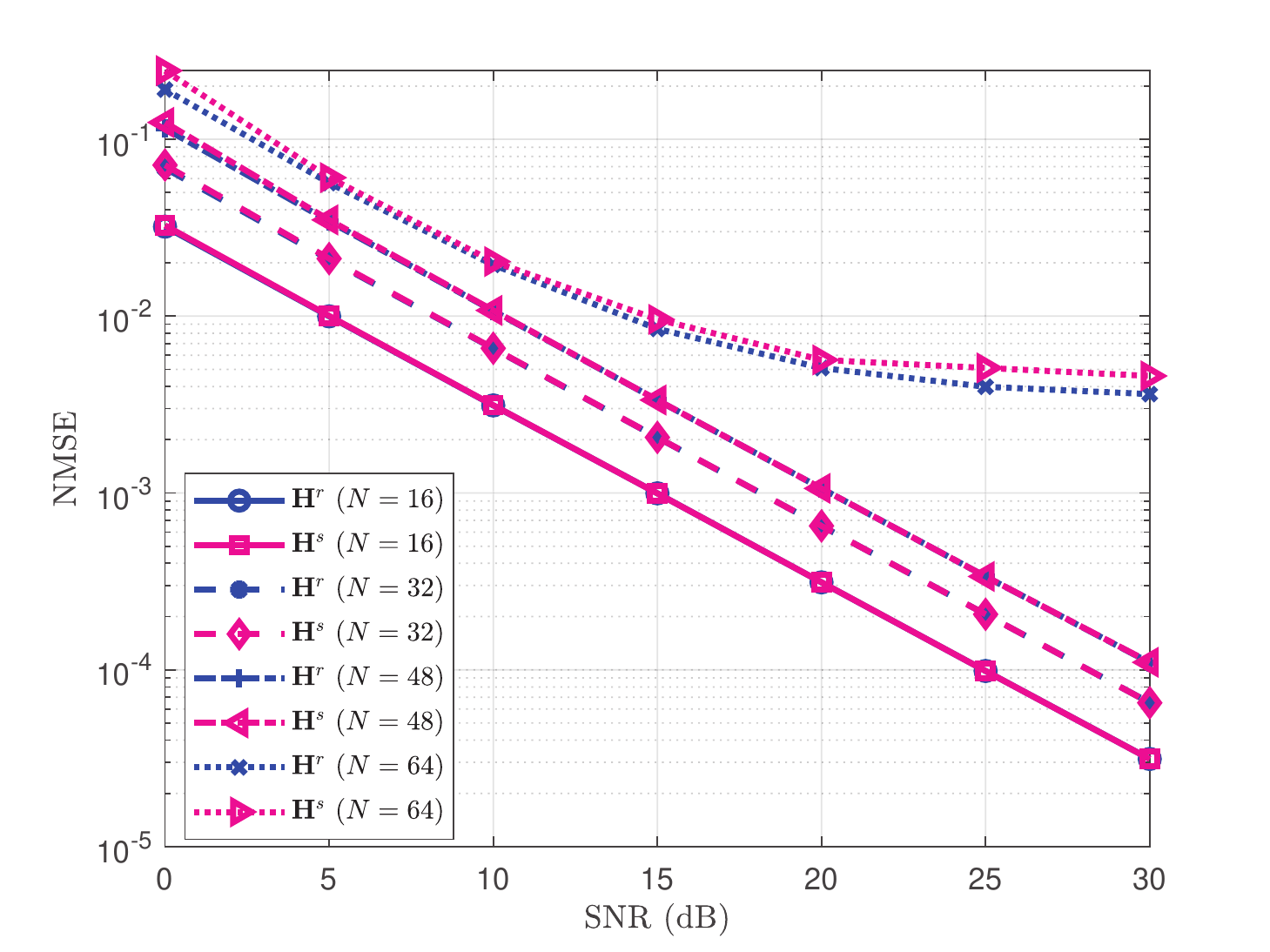} } \vspace{-2mm}
    		\caption{NMSE performance of the proposed ALS CE versus the SNRs in dB for $M=K=T=64$, $P=16$, and various values of $N$.}
    		\label{fig:N_Hr_Hs} \vspace{-6mm}
    	\end{center}
    \end{figure}

\begin{figure} \vspace{-1mm}
    	\begin{center}
    		\includegraphics[width=0.45\textwidth]{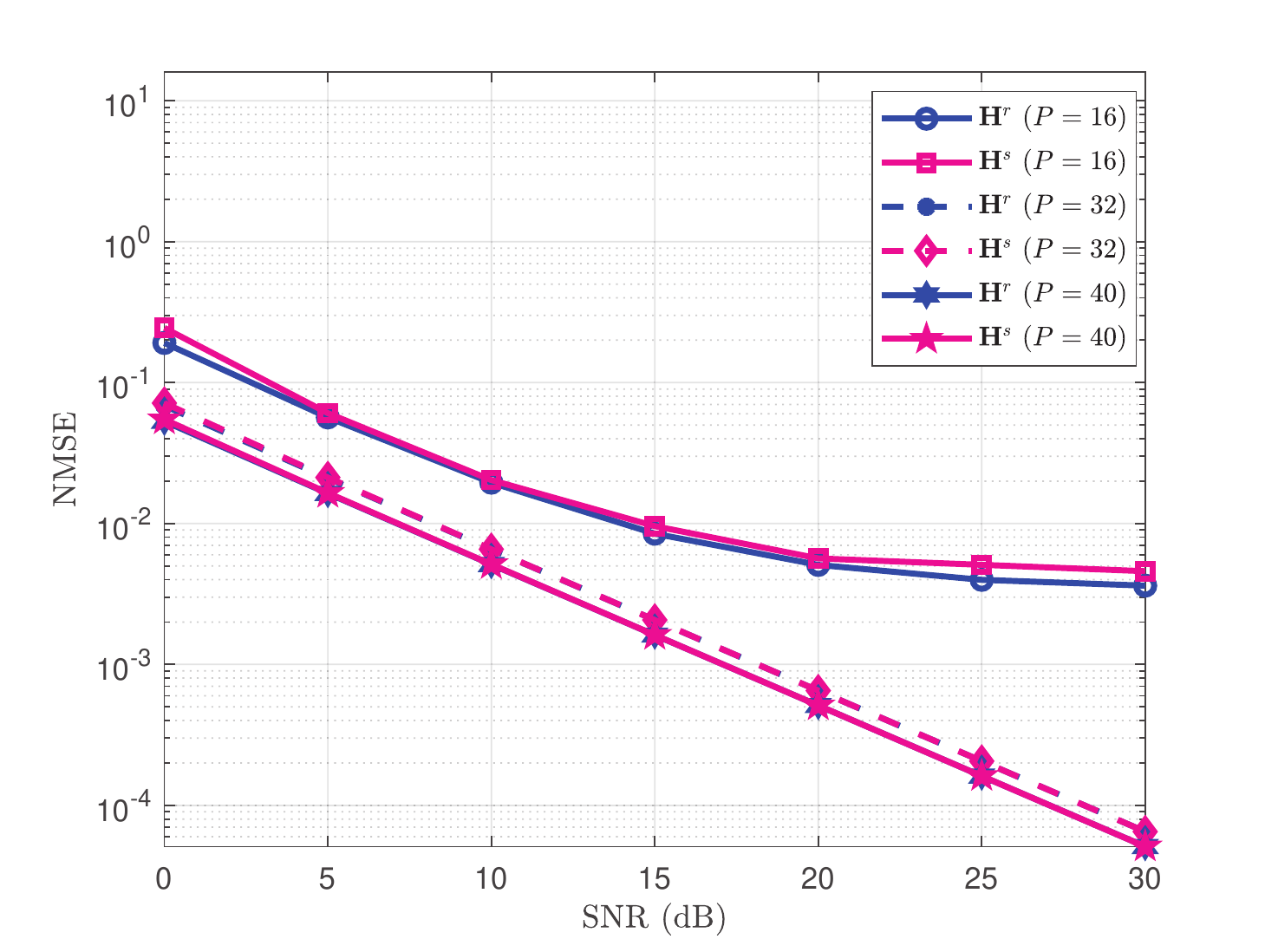}  \vspace{-2mm}
    		\caption{NMSE performance of the proposed ALS CE versus the SNRs in dB for $M=K=T=N=64$ and various values of $P$.}
    		\label{fig:P_Hr_Hs} \vspace{-6mm}
    	\end{center}
    \end{figure}

In Fig$.$~\ref{fig:N_Hr_Hs}, we have set the system parameters as $M=K=T=64$ and $P=16$, and simulated various values of $N$ for illustrating the NMSE performance of Alg. 1 as a function of the SNR. It is evident that there exists an increasing performance loss when increasing the number $N$ of RIS unit elements. A large $N$ results in a large number of rows and columns for $\mathbf{H}^s$ and $\mathbf{H}^r$, respectively, which requires more training pilots and larger computational complexity for ALS-based estimation to reduce the NMSE.
The impact of the number $P$ for the RIS training phase configurations in the NMSE performance is investigated in Fig$.$~\ref{fig:P_Hr_Hs}, where we compare the NMSE performance versus the SNRs for the setting $M=K=T=N=64$. It is shown that the increasing $P$ improves NMSE; for example, $P=40$ yields the best performance in the figure. This happens because when $P$ increases, the training set will increases and this is beneficial to CE. However, the speed of the performance improvement between adjacent NMSE curves becomes slower when $P>32$. It is also noted that the larger $P$ results in higher computational complexity and $P$ should be set less than $64$ according to the feasibility conditions of Lemma 1.
\subsection{CRB of the Proposed ALS CE}
\begin{figure} \vspace{-1mm}
		\begin{center}
			\includegraphics[width=0.45\textwidth]{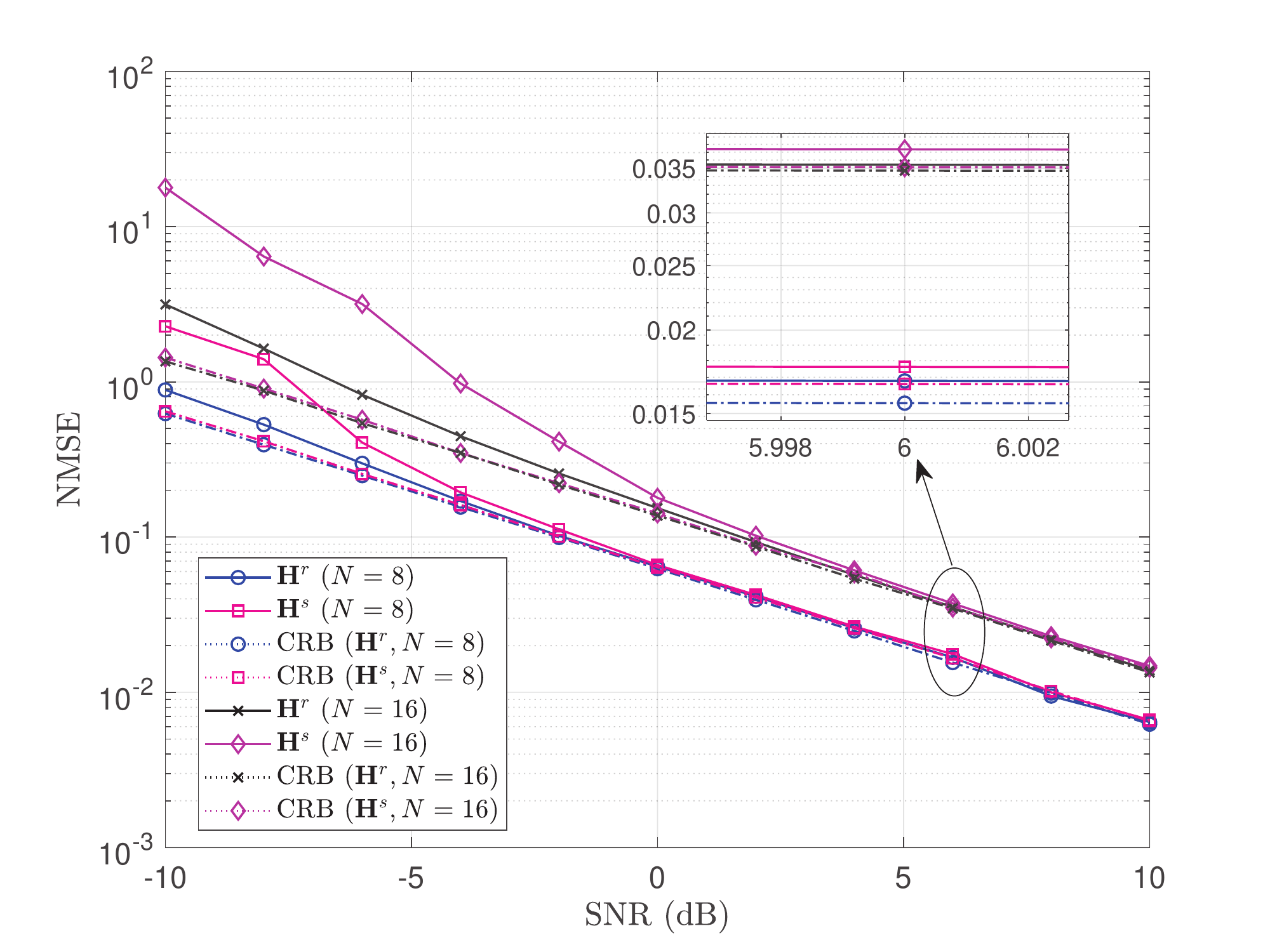}  \vspace{-2mm}
			\caption{CRBs of the proposed ALS CE versus the SNRs in dB for $M=K=T=32$, $P=8$, and various values of $N$.}
			\label{fig:crb_n} \vspace{-6mm}
		\end{center}
	\end{figure}

Fig.~\ref{fig:crb_n} contains the performance comparison between the CRBs and the NMSE performance of the proposed ALS Alg. 1 for the parameter setting $M=K=T=32$, $P=8$ and $N=\{8,16\}$. As shown, with the increase of $N$, the unknown variables increase and the estimated channel matrices are more complex, thus, the CE performance degrades. However, the performance gap reduces with the increase of the SNR, and the estimation performance achieves CRBs when SNR$\geq 2$ dB, i.e., the estimations of $\mathbf{H}^s$ and $\mathbf{H}^r$ achieve their respective CRBs. Overall, Fig.~\ref{fig:crb_n} proves that the proposed algorithm works well in different scenarios, and also achieves CRBs especially at the larger SNR regime.

  \subsection{Downlink Sum Rate}
\begin{figure} \vspace{-1mm}
		\begin{center}
			\includegraphics[width=0.45\textwidth]{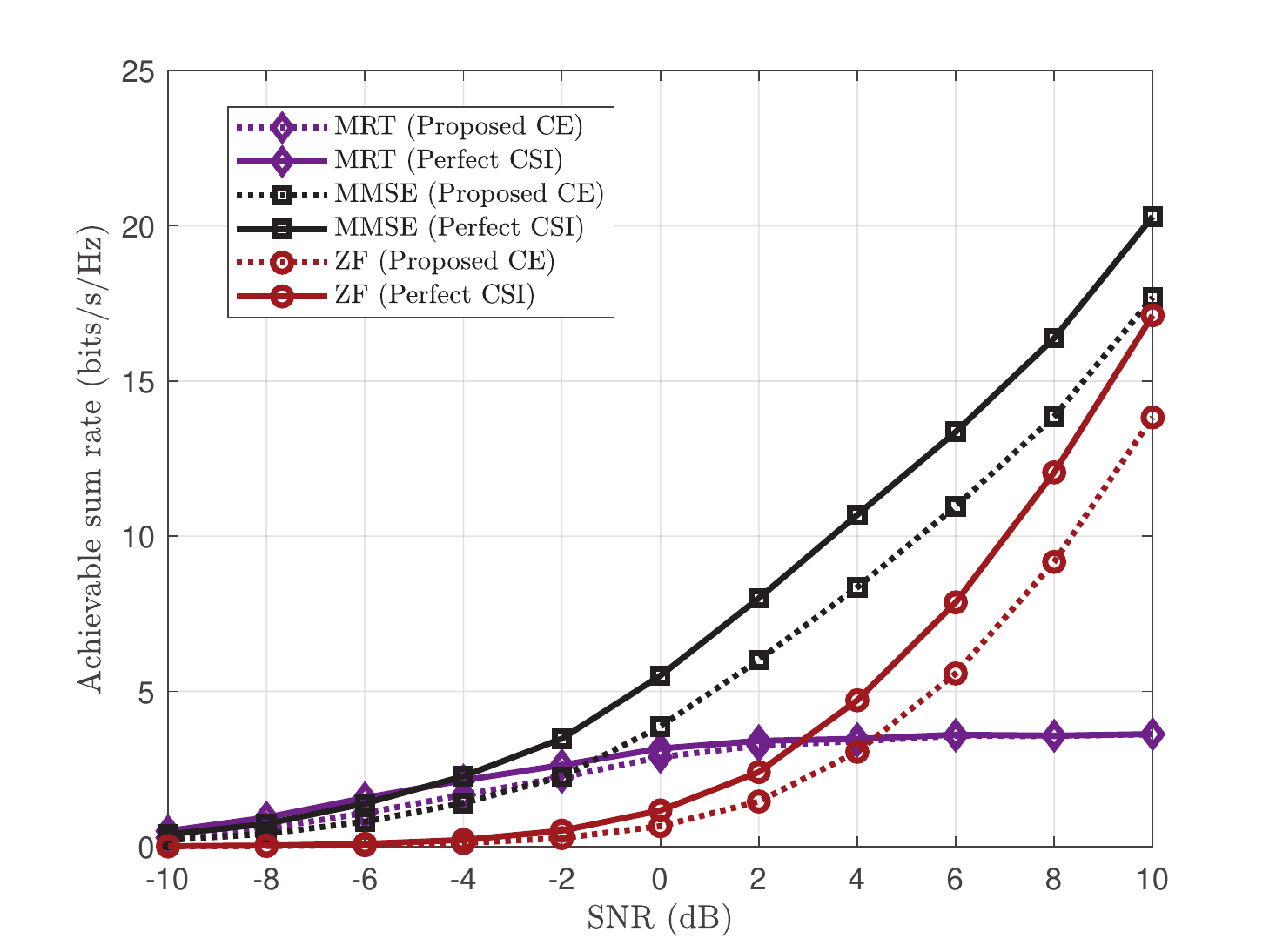}  \vspace{-2mm}
			\caption{Achievable sum rate of the proposed ALS CE with the MRT, MMSE, and ZF precoding schemes versus the SNRs in dB under the setting $M=6, K=T=N=P=4$.}
			\label{fig:sum_rate_all4} \vspace{-6mm}
		\end{center}
	\end{figure}

	The achievable sum rate with the proposed ALS CE algorithm using the MRT, MMSE, and ZF precoding schemes at BS is depicted in Fig.~\ref{fig:sum_rate_all4} for $M=6, K=T=N=P=4$ as a function of the operating SNR. It is shown that, as the SNR increases, the sum rate improves. We observe that MMSE beamforming performs the best in the entire SNR range, and MRT is close to MMSE at low SNRs, while ZF is close to MMSE at high SNRs. The achievable rate of ZF increases rapidly with the increase of SNRs while for MRT it increases slowly.  In addition, the gap between the proposed CE algorithms and the case of perfect CE with the MRT precoding scheme becomes narrower as SNRs increase, and the gap will be reduced as zero when  SNR $\geq 4$ dB.
	
\begin{figure} \vspace{-1mm}
		\begin{center}
			\includegraphics[width=0.45\textwidth]{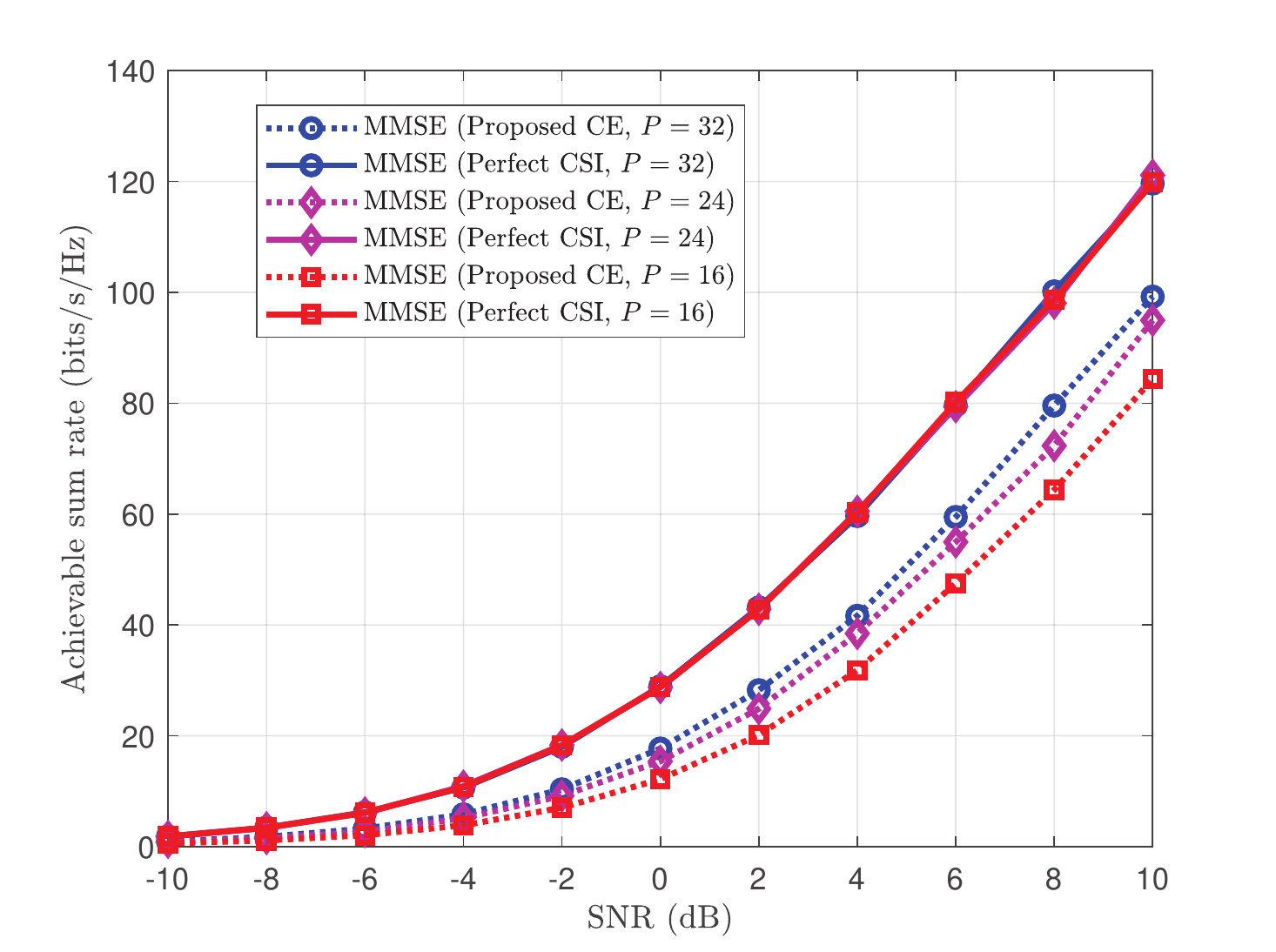}  \vspace{-2mm}
			\caption{Achievable sum rate of the proposed ALS CE with MMSE precoding versus the SNRs in dB for $M=K=T=N=32$ and various values of $P$.}
			\label{fig:sum_rate_p_mmse} \vspace{-6mm}
		\end{center}
	\end{figure}

\begin{figure} \vspace{-1mm}
		\begin{center}
			\includegraphics[width=0.45\textwidth]{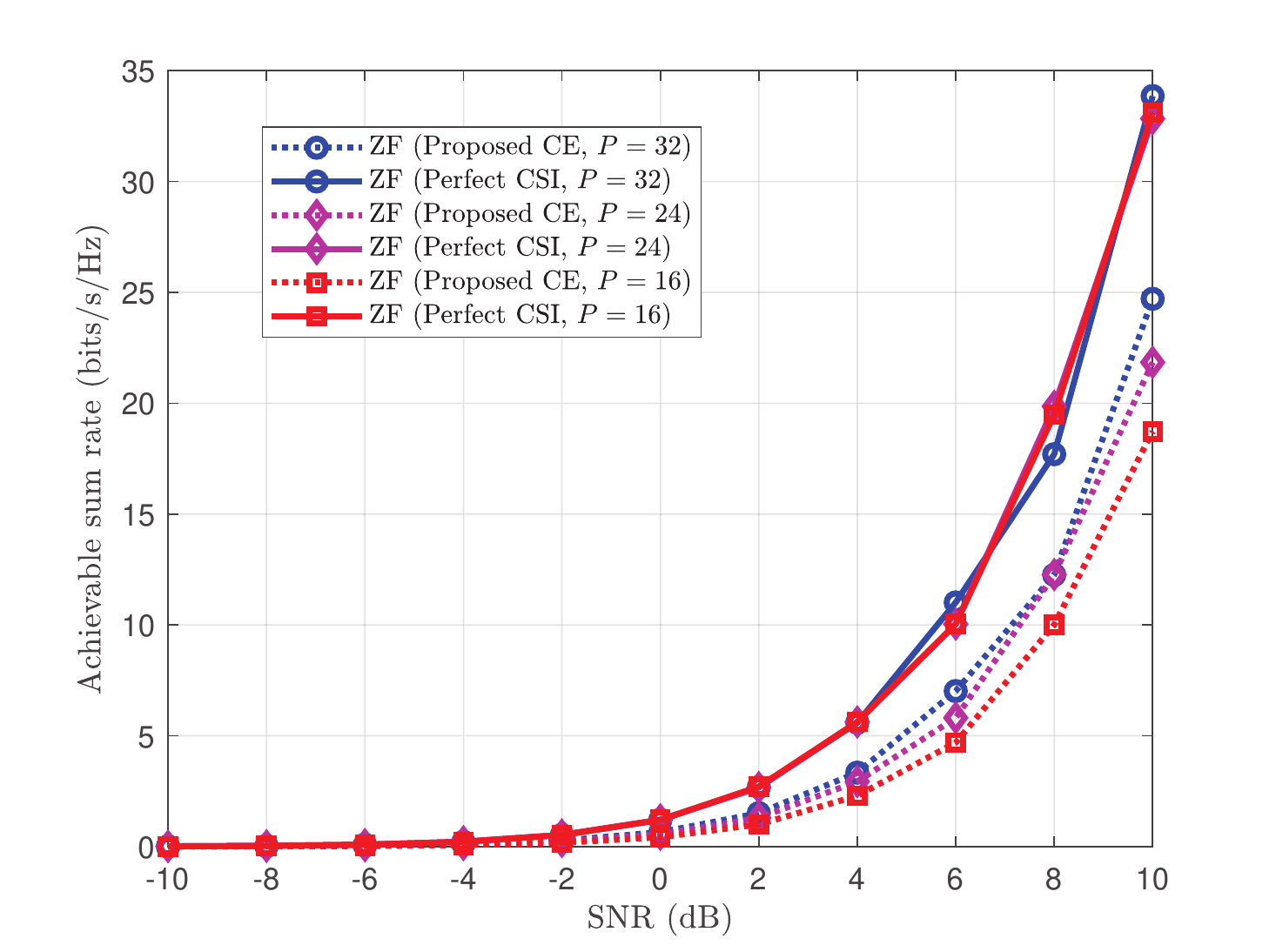}  \vspace{-2mm}
			\caption{Achievable sum rate the proposed ALS CE with ZF precoding versus the SNRs in dB for $M=K=T=N=32$ and various values of $P$.}
			\label{fig:sum_rate_p_zf} \vspace{-6mm}
		\end{center}
	\end{figure}

\begin{figure} \vspace{-1mm}
		\begin{center}
			\includegraphics[width=0.45\textwidth]{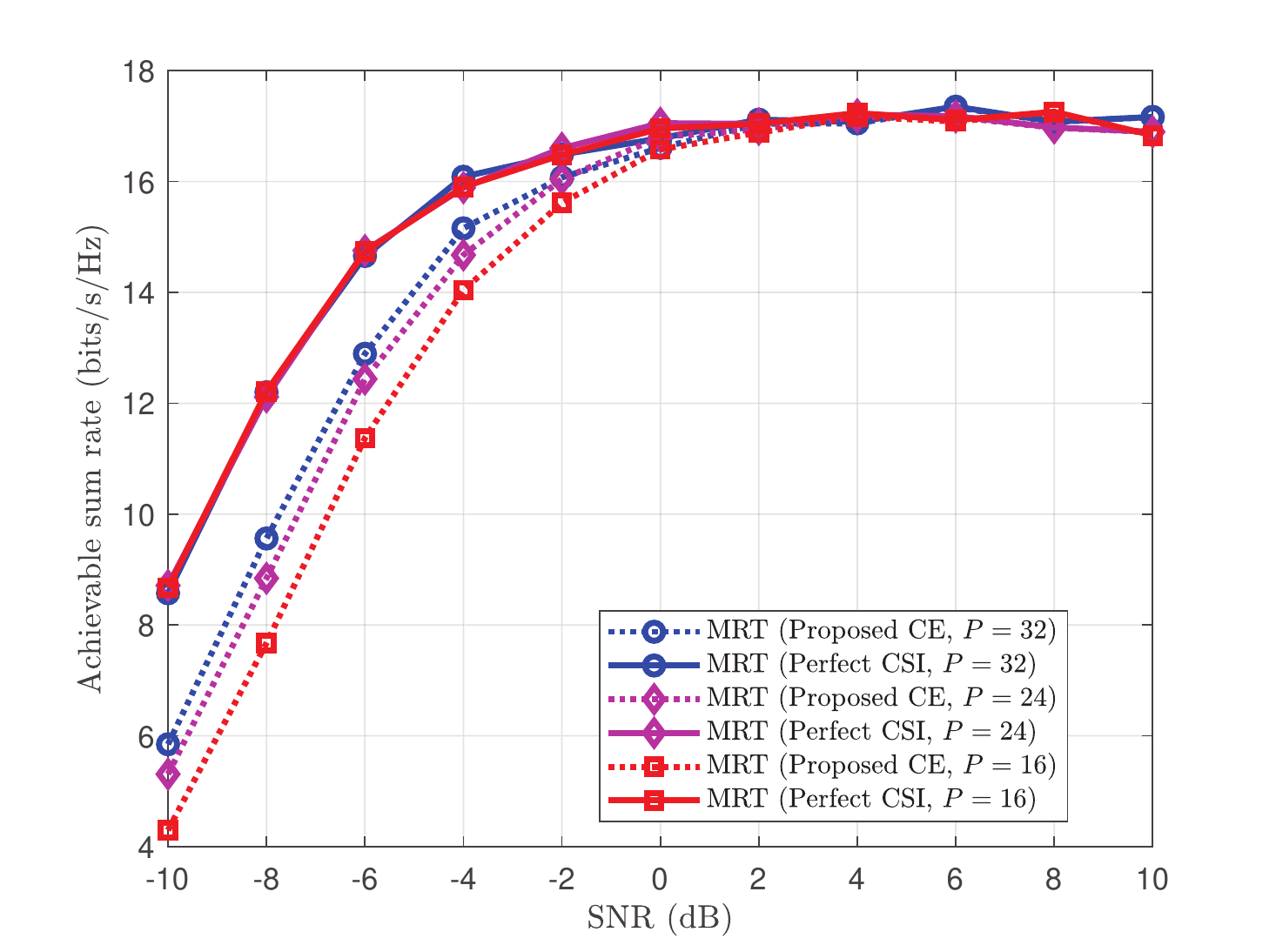}  \vspace{-2mm}
			\caption{Achievable sum rate the proposed ALS CE with MRT precoding versus the SNRs in dB for $M=K=T=N=32$ and various values of $P$.}
			\label{fig:sum_rate_p_mrt} \vspace{-6mm}
		\end{center}
	\end{figure}

	Fig.~\ref{fig:sum_rate_p_mmse} shows the sum rate of the proposed ALS algorithm for different values of $P$ using the ZF precoding scheme under the setting $M=K=T=N=32$ and $P=\{16,24,32\}$. It can be seen easily that with the increase of $P$ will result in the improvement of the achievable sum rate under considered all cases. This is because that a larger P means more training pilots and thus CE performance improves, which is also shown in Fig.~\ref{fig:P_Hr_Hs}. However, the performance gaps between the curves for different $P$ become narrower when the SNR increases. The similar results are also demonstrated in Fig.~\ref{fig:sum_rate_p_zf} and Fig.~\ref{fig:sum_rate_p_mrt} for scenarios where MMSE and MRT precoding schemes are adopted at the BS respectively.
	
\begin{figure} \vspace{-1mm}
		\begin{center}
			\includegraphics[width=0.45\textwidth]{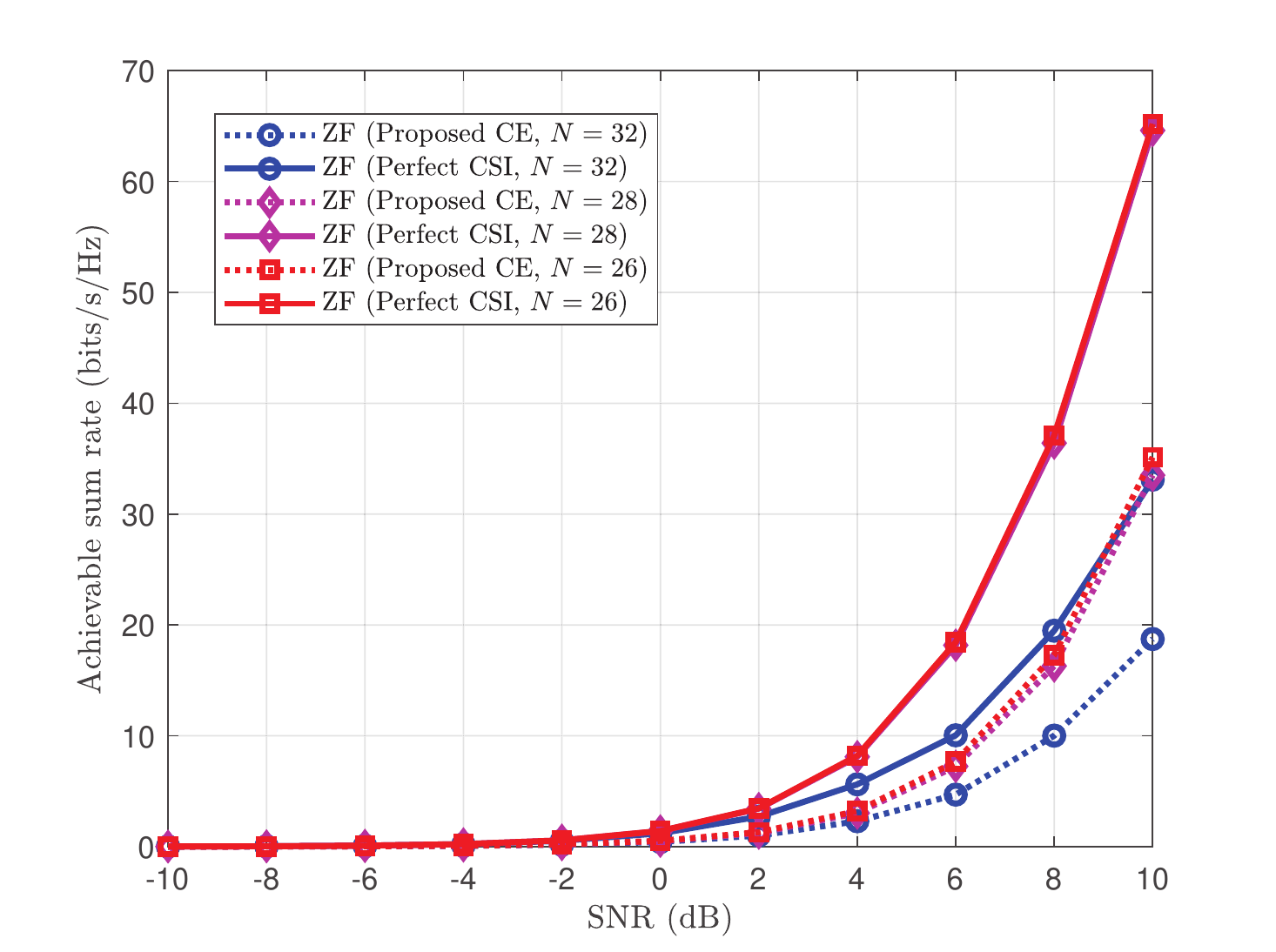}  \vspace{-2mm}
			\caption{Achievable sum rate of the the proposed ALS CE with ZF precoding versus the SNRs in dB for $M=K=T=32,P=16$ and various values of $N$.}
			\label{fig:sum_rate_n_zf} \vspace{-6mm}
		\end{center}
	\end{figure}

\begin{figure} \vspace{-1mm}
		\begin{center}
			\includegraphics[width=0.45\textwidth]{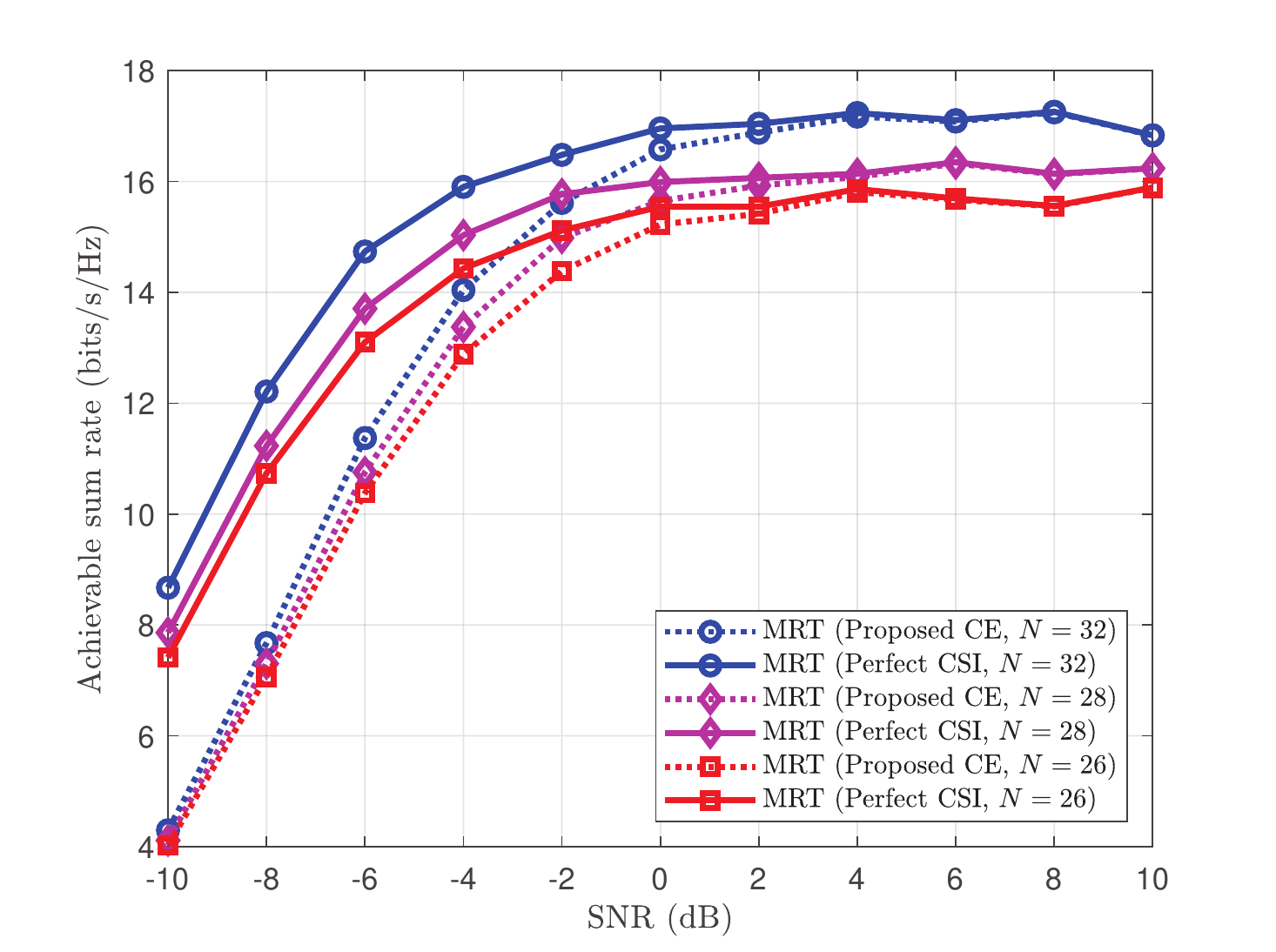}  \vspace{-2mm}
			\caption{Achievable sum rate of the proposed ALS CE with MRT precoding versus the SNR in dB for $M=K=T=32,P=16$ and various values of $N$.}
			\label{fig:sum_rate_n_mrt} \vspace{-6mm}
		\end{center}
	\end{figure}

	Finally, in Fig.~\ref{fig:sum_rate_n_zf} and Fig.~\ref{fig:sum_rate_n_mrt}, we plot the sum rate of the proposed ALS algorithm for $M=K=T=32$ and $P=16$, and different values of $N$ using the ZF and MRT precoding schemes, respectively. It is shown from Fig.~\ref{fig:sum_rate_n_zf} that a larger $N$ value leads to a degraded sum rate performance in the high SNR regime. The larger $N$ results the less allocated power at each user, thus, the sum rate degrades as well. Unlike the curves in Fig.~\ref{fig:sum_rate_n_zf}, although the sum rate with the MRT scheme in Fig$.$~\ref{fig:sum_rate_n_mrt} has the improved performance with larger $N$, the sum rate reaches the floor when the SNR $\geq 0$ dB due to the useless suppression of inter-user interference. In addition, it is also shown that for the different values of $N$, the performance gap between the curves becomes narrower with increasing SNR. Especially for SNR $\geq 4$ dB, and the sum rate of the proposed algorithm is close to that of the perfect CE case in Fig$.$~\ref{fig:sum_rate_n_mrt}.

\section{Conclusion}\label{sec:conclusion}
In this paper, we presented two iterative CE techniques, one based on ALS and another one based on VAMP, for RIS-empowered multi-user MISO uplink communication systems that capitalize on the PARAFAC decomposition of the received signal model. We analytically investigated the feasibility conditions and the computational complexity of both proposed algorithms, and we derived the CRB for the ALS-based CE.    Analysis shows that VAMP-based algorithm has the lower complexity than ALS-based algorithm. However, the proposed algorithms also have the limitations, e.g., the element number of RIS needs to meet the constraints of Lemma 1, and they also need to reduce the performance to remove the estimation ambiguity.   Our simulation results verified that both proposed algorithms exhibit similar performance and both outperform the benchmarked state-of-the-art LSKRF scheme, while being quite robust to the induced noise. Moreover, it was also shown that the numbers of RIS unit elements and training symbols exert a significant effect on the performance of the proposed algorithms. We also studied the achievable sum rate performance with estimated channels based on our proposed algorithms  considering the MRT, MMSE, and ZF precoding schemes.

\ifCLASSOPTIONcaptionsoff
\newpage
\fi

	
\bibliographystyle{IEEEbib}
\bibliography{strings}
 

\end{document}